\pdfoutput=1
\documentclass{article}

\usepackage[english]{babel}
\usepackage[utf8]{inputenc}
\usepackage{color}

\usepackage{graphicx} 
\usepackage{amsmath} 
\usepackage{amssymb} 
\usepackage{amsthm} 
\usepackage{tabularx} 
\usepackage{siunitx} 
\usepackage{algorithm} 
\usepackage[noend]{algpseudocode} 
\usepackage{caption} 
\usepackage{subcaption} 
\usepackage{filecontents} 

\usepackage{authblk}

\everymath{\displaystyle}
\graphicspath{{img/}}

\title{ A 2D front-tracking Lagrangian model for the modeling of anisotropic grain growth }
\author[1]{Sebastian~Florez\thanks{corresponding author}}
\author[1]{Julien~Fausty}
\author[1]{Karen~Alvarado}
\author[1]{Brayan~Murgas}
\author[1]{Marc~Bernacki}

\affil[1]{Mines-ParisTech, PSL-Research University, CEMEF – Centre de mise en forme des mat\'{e}riaux, CNRS UMR 7635, CS 10207 rue Claude Daunesse, 06904 Sophia Antipolis Cedex, France}%

\begin{document}
\maketitle

\section*{abstract}

Grain growth is a well-known and complex phenomenon occurring during annealing of all polycrystalline materials. Its numerical modeling is a complex task when anisotropy sources such as grain orientation and grain boundary inclination have to be taken into account. This article presents the application of the front-tracking methodology \emph{ToRealMotion} introduced in previous works, to the context of anisotropic grain boundary motion at the mesoscopic scale. The new formulation of boundary migration can take into account any source of anisotropy both at grain boundaries as well as at multiple junctions (MJs) (intersection point of three or more grain boundaries). Special attention is given to the decomposition of high-order MJs for which an algorithm is proposed based on local grain boundary energy minimisation. Numerical tests are provided using highly heterogeneous configurations, and comparisons with a recently developed Finite-Element Level-Set (FE-LS) approach are given. Finally, the computational performance of the model will be studied comparing the CPU-times obtained with the same model but in an isotropic context.

\section{Introduction}\label{sec:introduction}

Grain growth phenomenon in polycrystals has been studied for many decades, both from an experimental and numerical point of view \cite{Humphreys2017}. The majority of experimental observations at this scale suggest that the migration of boundaries is, in general, a strongly heterogeneous phenomenon involving complex dynamics and topological transformations of the grain boundary (GB) network. However, in the literature, it is frequently accepted that the microstructure of given materials behave homogeneously enough to ignore their heterogeneities when considering polycrystal modelling. This hypothesis is used in numerical environments to propose Full-Field (FF) models of microstructural evolutions, using homogeneous values in space of the grain boundary energy $\gamma$ and mobility $\mu$, e.g., isotropic grain growth (GG).\\
If this hypothesis remains acceptable when low levels of anisotropy are involved, it constitutes, however, a strong approximation when a strong texture with particular $\gamma$ values are involved, or when special GBs (e.g., twin boundaries) are present \cite{watanabe2011grain}.\\

Commonly in the literature \cite{Humphreys2017}, the source of GB anisotropy, i.e. the reduced mobility defined as the $\mu\gamma$ product, is considered as a function of the crystallographic misorientation and of the inclination of the interface. Typically,  the misorientation $M_{lw}$ between two adjacent grains $l$ and $w$, is computed using the three Euler angles $(\varphi_{e1}, \Phi, \varphi_{e2})$ of these grains and the inclination is considered through the \emph{local} normal vector $\vec{n}$ of the corresponding GB. This gives a system with a total of 5 degrees of freedom (DOF) defining GB properties. These kinds of systems at the polycrystal scale need to be modelled through the use of a numerical approach \emph{able to take into account} this kind of data set. As such, in the same manner as in \cite{Fausty2020a, Fausty2020b, Murgas2021}, here we will differentiate three kinds of numerical models: isotropic, heterogeneous, and anisotropic models. \textbf{Isotropic} models consider constant GB properties in their formulation. On the contrary, \textbf{Anisotropic} models, are those using a formulation where, any assumption regarding the invariability of these quantities in space is discarded, being able to use properties dependent on the tuple $(M_{lw}, \vec{n})$ (i.e. $X_{(M_{lw}, \vec{n})}$ where $X$ is either $\gamma$ or $\mu$). Of course, anisotropic models are much more complex than those using an isotropic hypothesis, since, in this context, special attention must be given, for example, to the meaning of the surface tension component of interfaces, as one must be aware that \emph{torque terms}, derived from the variation of the GB energy $\gamma$ on the parametric space of a surface may appear \cite{DeWit1959}. As such, deriving a mathematical model ready to use an anisotropic set of GBs properties is a complex task, and historically, authors in the literature have proposed alternatives: heterogeneous models. \textbf{Heterogeneous} models consider within their formulation the existence of a variation of properties, only in function of $M_{lw}$ ($X_{(M_{lw})}$), neglecting its dependence on $\vec{n}$. In this context, each GB is given \emph{homogenised} intrinsic properties (constant in its parametric space), but different from the ones of other GBs. i.e., GB properties only change at multiple junctions (MJ) (or multiple lines in 3D) when crossing from one GB to another.\\

Several approaches have been proposed in the literature to model heterogeneous/anisotropic GG. Beginning with the Monte Carlo and extending to Phase-Field, Level-Set and Vertex approaches, heterogeneous ($X_{(M_{lw})}$) \cite{Grest1985, Fausty2018, Fausty2020, Miyoshi2017}, and anisotropic ($X_{(M_{lw}, \vec{n})}$) \cite{Fausty2020a, Hallberg2019, BarralesMora2010,Fausty2020b,Murgas2021} models have been proposed. However, all these methods are constrained by different reasons each, typically: i. the use of regular grids \cite{Elsey2013, Miessen2017} (which can lead to difficulties to model large deformation), ii. high computational cost \cite{Fausty2018,Fausty2020b,Murgas2021}, and iii. the no-discretization of grain interiors \cite{Kawasaki1989, BarralesMora2010} (which can lead to difficulties when intragranular phenomena are of interests). Additionally to these aspects, in the context of anisotropic boundary properties modelled using Phase-Field models, although being an appropriate numerical environment, showing interesting results in this context, one should be aware of inherent numerical instabilities, especially for high heterogeneous/anisotropic systems \cite{Miyoshi2017, Chang2017}. Finally, in anisotropic models, the GB energy dependence on the inclination is classically defined without inquiring if additional torque terms in solved equations are needed with the notable exception of the vertex approaches \cite{BarralesMora2010,Kawasaki1989}. \\

As an alternative to model microstructural evolutions with anisotropic GB properties, we propose the TRM model presented in \cite{Florez2020b, Florez2020c, Florez2020d}: this article will present the needed implementations in order to model fully anisotropic grain properties with the TRM model. Special attention will be given to the development of a robust high order Multiple Junction (MJ) decomposition algorithm and to the reformulation of the velocity equation at triple junctions extending the methodology presented in \cite{Kawasaki1989} to an anisotropic context, using the notions used in \cite{BarralesMora2010} for its discrete formulation. Finally, the TRM model will be tested in multiple heterogeneous environments identical to the ones presented in \cite{Fausty2018, Fausty2020} while the numerical tests in a fully anisotropic environment will be discussed in a forthcoming publication \cite{Florez2021Statistical}.

\section{Numerical method}\label{sec:numericalmethod_5}

This section will introduce the TRM model's necessary concepts and new implementations to model GBM using anisotropic GB properties. The topological changes that may occur in this context have the same level of complexity as the ones produced under the influence of stored energy, presented in a previous work \cite{Florez2020d}. Additionally, in \cite{Florez2020b}, the decomposition of high-order multiple junctions (MJs) was simplified for isotropic GB properties. A more developed algorithm is needed to obtain valid predictions in an anisotropic context. This section will cover these notions.\\

Hereafter, we will consider $\gamma$ as a function of $\left(M_{lw}, \vec{n}\right)$, while the mobility term $\mu$ will be considered constant in space. Before considering a misorientation in the computation of grain boundary properties, each grain requires an orientation. In this work, these orientations are generated at random using a uniform distribution for each of Euler's angles ($\varphi_1$, $\Phi$, $\varphi_2$). Fig.~\ref{fig:2DGGHetero_5000_InitialState}(bottom-right) gives un example of the disorientation angle distribution obtained with this approach and compares it to the Mackenzie distribution \cite{Mackenzie1957} for disorientation angles in a cubic sample.\\

\subsection{Grain boundary motion by capillarity: Anisotropic context for the TRM model}\label{sec:VeloAniso}

In \cite{BarralesMora2010} a formulation for the computation of the velocity of GBs and triple junctions using anisotropic GB properties was proposed in the context of the Vertex model. This formulation uses the tensile character of the capillarity forces exerted at every node based on a discrete analysis, similar to the one used in \cite{Florez2020d} for the computation of a velocity from a stored energy field at triple junctions. The model in \cite{BarralesMora2010} writes for the velocity at MJs: 

\begin{equation}
\label{Eq:BarralesVelocity}
\centering
\vec{v_c}_i=\mu_i\left(\sum_j{\frac{\gamma_{ij} \vec{t}_{ij} + \tau_{ij}\vec{n}_{ij}}{ |\overline{N_iN_j}|}}\right),
\end{equation}

where the index $i$ denotes the node representing the MJ $N_i$ and $j$ their connection to node $N_j$, $\mu_i$ is the mobility of node $N_i$, $\gamma_{ij}$, $\vec{t}_{ij}$ and $\vec{n}_{ij}$ are respectively the boundary energy, the unit tangent vector and the normal of the segment $\overline{N_iN_j}$. Note that $\gamma_{ij}=\gamma_{ji}$ but $\vec{t}_{ij}=-\vec{t}_{ji}$ and the direction of the normal $\vec{n}_{ij}$ is arbitrary. Finally, note the apparition of the term $\tau_{ij}$, which corresponds to the torque experienced by the segment $\overline{N_iN_j}$ due to the change of the GB energy given by its dependence on the inclination angle $\omega$ \cite{DeWit1959}. This torque term is defined as follows:

\begin{equation}
\label{Eq:TorqueTerm}
\centering
\tau_{ij}=-\frac{d\gamma}{d\omega_{ij}}.
\end{equation}

In \cite{Kawasaki1989}, three formulations were given for the computation of the velocity at MJs in the context of isotropic GB properties, from which we have used the so-called model {\bf II} to find our velocity at MJs in previous works \cite{Florez2020b,Florez2020d}. This formulation can be rewritten in the context of heterogeneous grain boundary properties (hence, in the absence of torque terms) and for MJs of arbitrary order, in a very similar way as in Eq.~\ref{Eq:BarralesVelocity}:

\begin{equation}
\label{Eq:Model2KawaVelocity}
\centering
\vec{v_c}_i=\mu_i\left({\frac{\sum_j\gamma_{ij} \vec{t}_{ij}}{ c^{-1}  \sum_j |\overline{N_iN_j}|}}\right),
\end{equation}

where $c$ is the number of connections of the MJ, and where the only difference with Eq.~\ref{Eq:BarralesVelocity} is that the terms in the numerator contribute all in the same amount to the summation, instead of being escalated each by the term $|\overline{N_iN_j}|^{-1}$. Indeed, in our experience, the homogenization of the contributions of the numerator term by the separated summation $\sum_j|\overline{N_iN_j}|$ has proven to be more stable than the one given in Eq.~\ref{Eq:BarralesVelocity}, especially when the value of any $|\overline{N_iN_j}|$ approaches to zero (or when $|\overline{N_iN_j}| << |\overline{N_iN_k}|$ for all $k\neq j$). For this reason, the use of Eq.~\ref{Eq:Model2KawaVelocity} is preferred here but maintaining the torque terms of Eq.~\ref{Eq:BarralesVelocity}:

\begin{equation}
\label{Eq:Model2KawaVelocitywithTorque}
\centering
\vec{v_c}_i=\mu_i\left({\frac{\sum_j\gamma_{ij} \vec{t}_{ij} + \tau_{ij}\vec{n}_{ij}}{ c^{-1}  \sum_j|\overline{N_iN_j}|}}\right).
\end{equation}

Finally, note that torque effects also need to be considered at GBs. For this purpose, the analytical model introduced in \cite{herring1999surface} for single surfaces can be used:

\begin{equation}
\label{Eq:Model2KawaHerring}
\centering
\vec{v_c}_i=\mu_i\left( -{\gamma_{i} + \frac{\partial^2 \gamma_{i}}{\partial n_{ix}^2}} \right)\kappa_i\vec{n}_i,
\end{equation}

where $n_{ix}$ is the projection of the \emph{variable} normal vector $\vec{n}$, onto the tangent vector to the interface at node $i$. In practice, we have found that applying Eq. \ref{Eq:Model2KawaHerring} might produce oscillatory effects on the computation of velocity $\vec{v_c}$ given by the second derivatives of $\gamma$. To avoid such instabilities, we use a combination of the discrete approach given in \cite{BarralesMora2010} and the standard approach of the TRM model:

\begin{equation}
\label{Eq:Model2KawaVelocityAndBarralesGB}
\centering
\vec{v_c}_i=\mu_i\left(-{\gamma_{i}\kappa_i\vec{n}_i + \frac{\sum_j \tau_{ij}\vec{n}_{ij}}{ c^{-1}\sum_j|\overline{N_iN_j}|}}\right),
\end{equation}

where the terms $\kappa_i$ and $\vec{n}_i$ are computed using the numerical approximation by splines at node $i$. Note that $\vec{n}_i\neq\vec{n}_{ij}$, as $\vec{n}_{ij}$ denotes the normal of the segment $\overline{N_iN_j}$ and $\vec{n}_i$ the normal to the numerical approximation at node $i$.\\

\subsection{Minimal-state energy of high-order MJs}\label{sec:highOrderMJSplitting}

\begin{figure}[!h]
\centering
\includegraphics[width=1.0\textwidth] {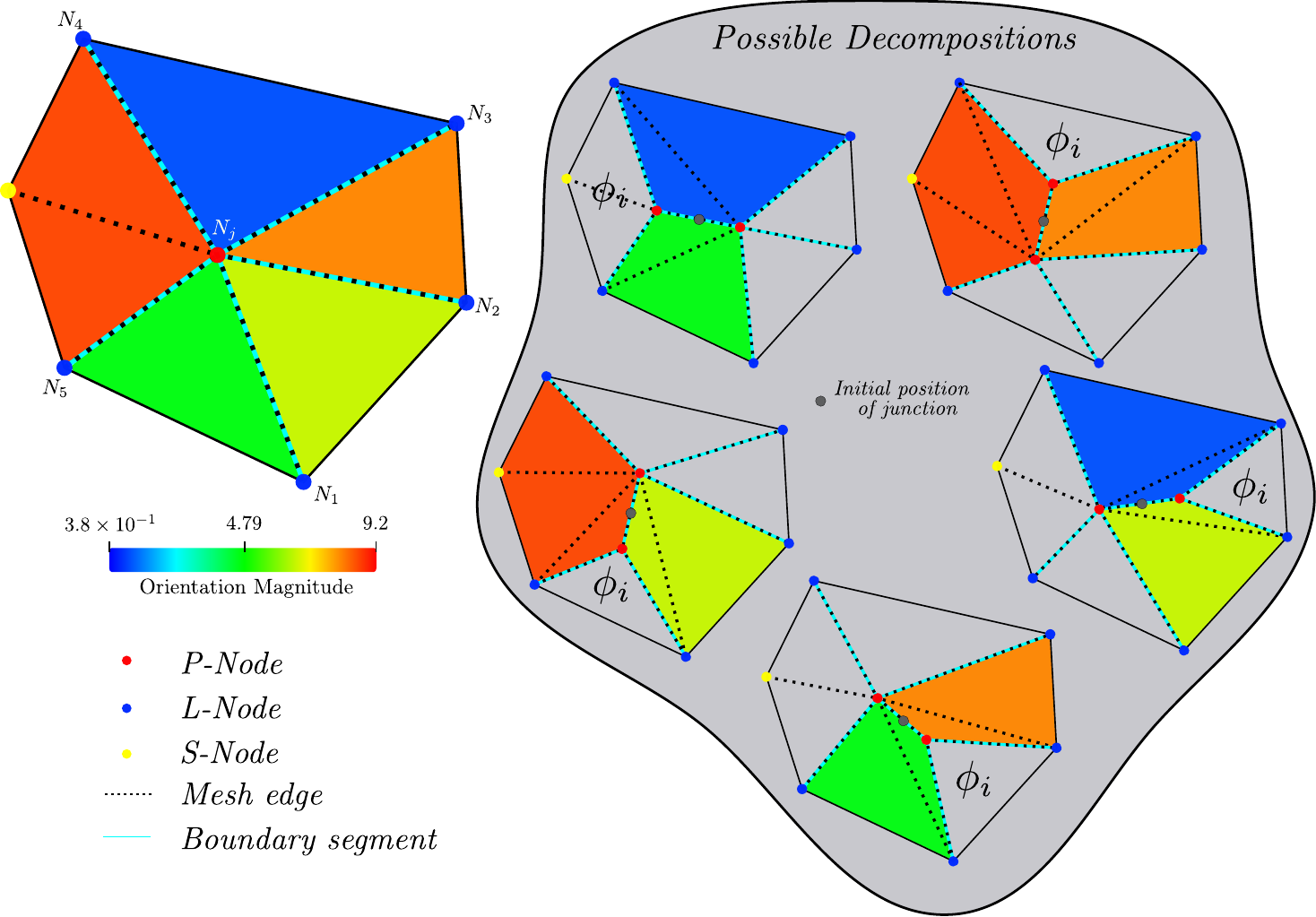}
\caption{Possible decompositions of a fifth-order MJ. Data regarding the orientation of each grain are given. One possible decomposition for every phase involved is depicted. The decomposition of a 5th order MJ results in a 4th order MJ and a  third-order MJ.}
\label{fig:PossibleDecompositions}
\end{figure}

This section provides an insight of the decomposition of high-order MJs when considering anisotropic GB properties.

\begin{figure}[!h]
\centering
\includegraphics[width=0.8\textwidth] {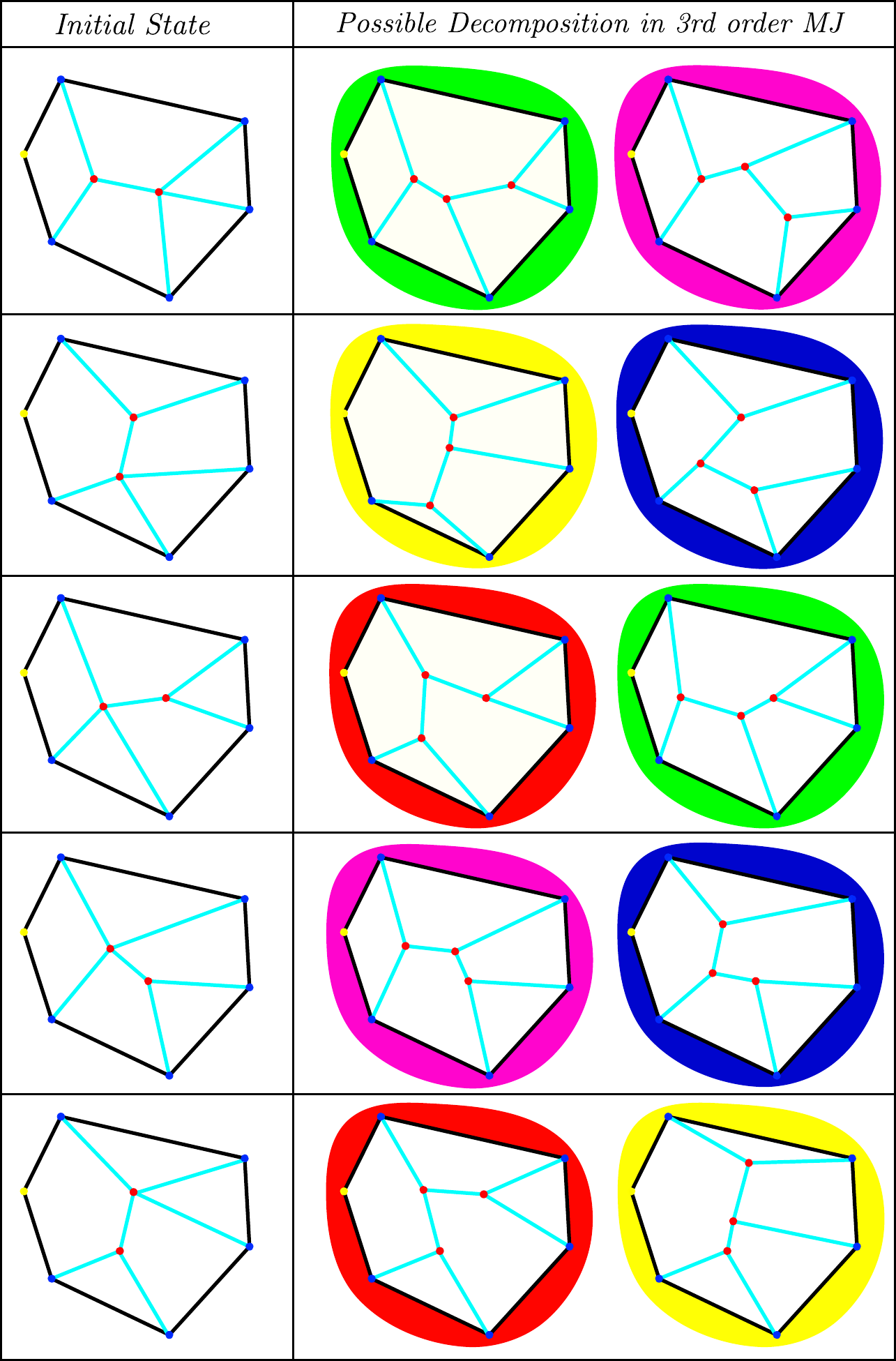}
\caption{Final possible decompositions of the MJ of Fig. \ref{fig:PossibleDecompositions} into MJ of 3rd order, the background colour of each final configuration matches similar configurations.}
\label{fig:PossibleDecompositions_Elemental}
\end{figure}

The main challenge here is to explore all possible configurations that may proceed after a decomposition process. The size of the possibilities set $P(z)$ is only dependent on the MJ's order $z$ to be decomposed. A 4th order MJ (i.e., four grain boundaries meeting in a point) can be decomposed only in two ways. However, the set of possibilities $P(z)$ grows much higher when the MJ's order increases. Consider the configuration given in Fig.~\ref{fig:PossibleDecompositions}, here we provide a fifth-order MJ, as well as the firsts five possible decompositions given by the separation of \emph{consecutive interface pairs}. Note, however, that each possibility regroups one fourth-order and one third-order MJ, from which the fourth-order one might decompose in two third-order MJs. In total, for a fifth-order MJ, five final possible decompositions are allowed when decomposing all MJ with $z>3$ ($P^3_{(5)}=5$ where the upper script means that all final MJs are $z=3$, see Fig.~\ref{fig:PossibleDecompositions_Elemental}). $P^3_{(z)}$ increase rapidly with the MJ's order $z$: for $z=(2, 3, 4, 5, 6, 7, 8,...)$,$P^3_{(z)}=(1,1,2,5,14,42,132,...)$. In general, the number of possible combinations in this context is given by the Catalan numbers \cite{Stanley1999} formula $C_{n}$:

\begin{equation}
\label{Eq:CatalanNumbers}
\centering
P^3_{(z)}=C_{(z-2)}=\frac{(2(z-2))!}{(z-1)!(z-2)!}
\end{equation}

Of course, the probability of encountering a MJ of order $z$ decreases as $z$ increases, as for a MJ of order $z$ to form, it would require that all $P_{(z-1)}$ possible decompositions were \emph{stable}. This notion of stability is related to the \emph{total} minimum energy state able to be reproduced for a given initial configuration. Note that this notion also suggests that one could obtain a \emph{total} minimal energy state for a MJ with $z>3$, in which case this MJ should not be decomposed \cite{BarralesMora2010}. As such, not only the configurations given by $P^3_{(z)}$ need to be considered, but also those in between (e.g. the ones given in Fig. \ref{fig:PossibleDecompositions}), hence giving $P_{(z)}>>P^3_{(z)}$.\\

We have simplified this aspect by accepting configurations presenting \emph{local} minimal energy states and by not testing all possible configurations $P_{(z)}$. Details regarding the decomposition algorithm are given in  Appendix \ref{apenDecompositionAlgorithm}.

\section{Numerical results}\label{sec:numericalresults_5}

In this section, the TRM model will be tested in a \emph{heterogeneous} context, meaning that the influence of the inclination angle $\omega$ over the value of $\gamma$ will be ignored. Then, $\gamma$ depends only on the disorientation angle $\theta$ and is equal for all segments defining the boundary between two given grains but different from all other boundaries. In such a context, the torque term $\tau$ is equal to 0 for all boundary segments, and the velocity of all nodes can be computed using Eq.~\ref{Eq:Model2KawaVelocity}.\\

All tests performed in this section have been inspired by the ones presented in \cite{Fausty2018, Fausty2020} in the same heterogeneous context. In \cite{Fausty2018}, the classical FE-LS formulation of \cite{Merriman1994, Bernacki2008, Scholtes2015} has been reformulated with the primary objective of taking into account the gradients terms produced by a variation of $\gamma$, that were otherwise neglected in a homogeneous context (where $\gamma$ is constant in $\Omega$). Given that this formulation considers all variational terms relevant in this context, it will be named hereafter the \emph{heterogeneous} FE-LS formulation. The numerical testing of this approach was divided into two parts: firstly, in \cite{Fausty2018}, the numerical analysis is focused on the evolution of multiple junctions as a means to test the \emph{heterogeneous} FE-LS formulation presented in the same publication. Secondly, in \cite{Fausty2020}, the same \emph{heterogeneous} FE-LS formulation was tested in the context of heterogeneous GG, using different formulations for the computation of the grain boundary energy $\gamma$, as a function of the disorientation angle. The approach used for the computation of the misorientation and disorientation angles can be found in Appendix \ref{apenMisorientComp}. \\

We reproduce these studies in the following with the TRM model:

\subsection{Triple junction test case}

\begin{figure}[!h]
\centering
\includegraphics[width=0.95\textwidth] {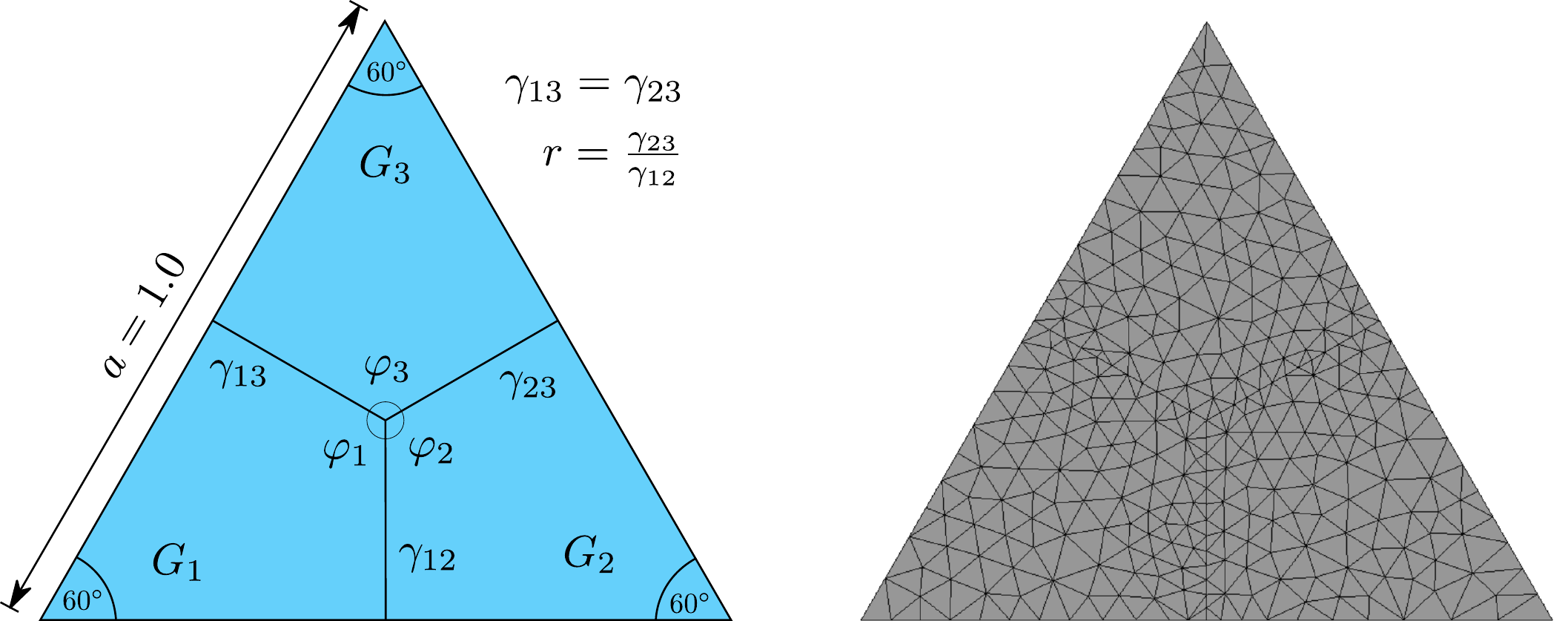}
\caption{Initial state for the triple junction test, three phases immersed in a domain in the shape of an equilateral triangle, the intersection of the domain with the interfaces is fixed (Dirichlet conditions $\vec{v}_i=0~\forall~N_i \in \partial \Omega$). a) Initial configuration and b) initial mesh.}
\label{fig:MultJunc_InitialState}
\end{figure}

The first test corresponds to an academic triple junction test. Here the motion of MJs is dictated by the GB energies of the interfaces meeting at the central node. Fig.~\ref{fig:MultJunc_InitialState}(left) illustrates this aspect, where the term $\gamma_{ij}$ denotes the GB energy between grains $G_i$ and $G_j$ and $\phi_i$ is the angle measured at the junction between the interfaces of grain $G_i$ and the other two grains. For this test, $\gamma_{13}=\gamma_{23}$, this will provoke a vertical movement of the junction for any value of $\gamma_{12}\neq\gamma_{23}$, until it arrives at its equilibrium position. As such, we will study the motion and the equilibrium of the junction in function of the ratio $r=\gamma_{23}/\gamma_{12}$. This test used dimensionless simulations, the value of the grain boundary energies $\gamma_{13}=\gamma_{23}=0.1$  and the mobility term $\mu=1$ were held constant, moreover, for practical reasons\footnote{For $r<1$, the MJ moves downwards, which when using Neuman type boundary conditions, induce a global movement of the interfaces in the same direction, and eventually leads to the contact of the junction with the base of the triangle. This behaviour is not wanted in this context.}, Dirichlet boundary conditions with $\vec{v}_i=0~\forall~N_i \in \partial \Omega$ are imposed, hence impeding the movement of the nodes at the intersection of the GBs and the edges of the triangular domain. Finally, the mesh size parameter $h_{trm}=0.006$ and the time step $dt=5e-5$ will be used for all tests. These values were selected correspondingly to the limit of the stability region of the TRM model when using piece-wise polynomials (splines) as a means to obtain values of curvature, and normal \cite{Florez2020b}. The initial mesh is illustrated in Fig.~\ref{fig:MultJunc_InitialState}(right). \\

While there is not an analytical formulation for the movement of the triple junction during its transient state in this context, triple junctions present stationary dihedral angles relying on the energies of the grain boundaries meeting at the junction \cite{herring1999surface}. In the absence of torque terms, i.e., when the energy of each interface is maintained constant, the dihedral angles $\phi_1$, $\phi_2$ and $\phi_3$ (see Fig.~\ref{fig:MultJunc_InitialState}) verify the Young's equilibrium, leading to the relation:


\begin{equation}
\label{Eq:YoungsEquilibriumAngles}
\centering
\frac{\sin{\phi_1}}{\gamma_{23}}=\frac{\sin{\phi_2}}{\gamma_{13}}=\frac{\sin{\phi_3}}{\gamma_{21}}.
\end{equation}

\begin{figure}[!h]
\centering
\includegraphics[width=1.0\textwidth] {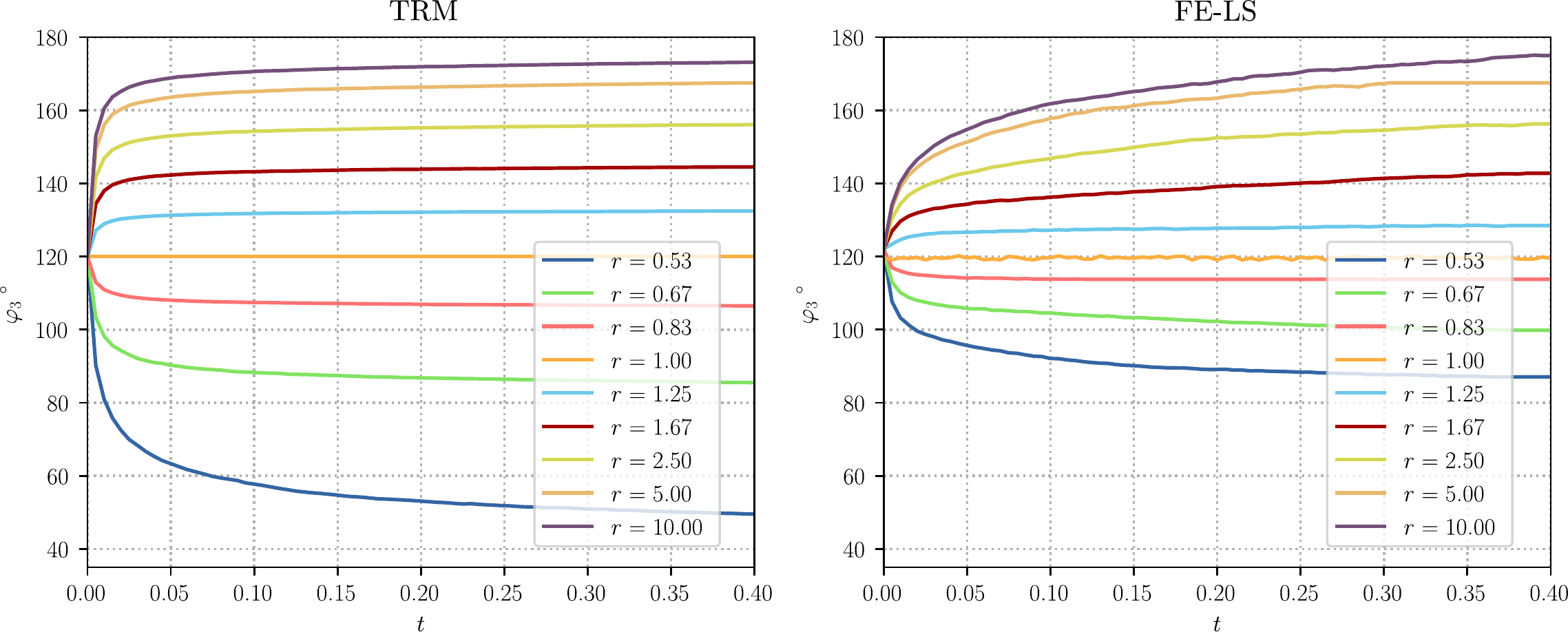}
\caption{Triple junction test case: Evolution of the $\varphi_3$ angle for different values of $r$, (left) TRM and (right) LS-FE models. The plotted data for the LS-FE model was taken from \cite{Fausty2018}.}
\label{fig:TriplePointAngleEvolution}
\end{figure}

\begin{figure}[!h]
\centering
\includegraphics[width=1.0\textwidth] {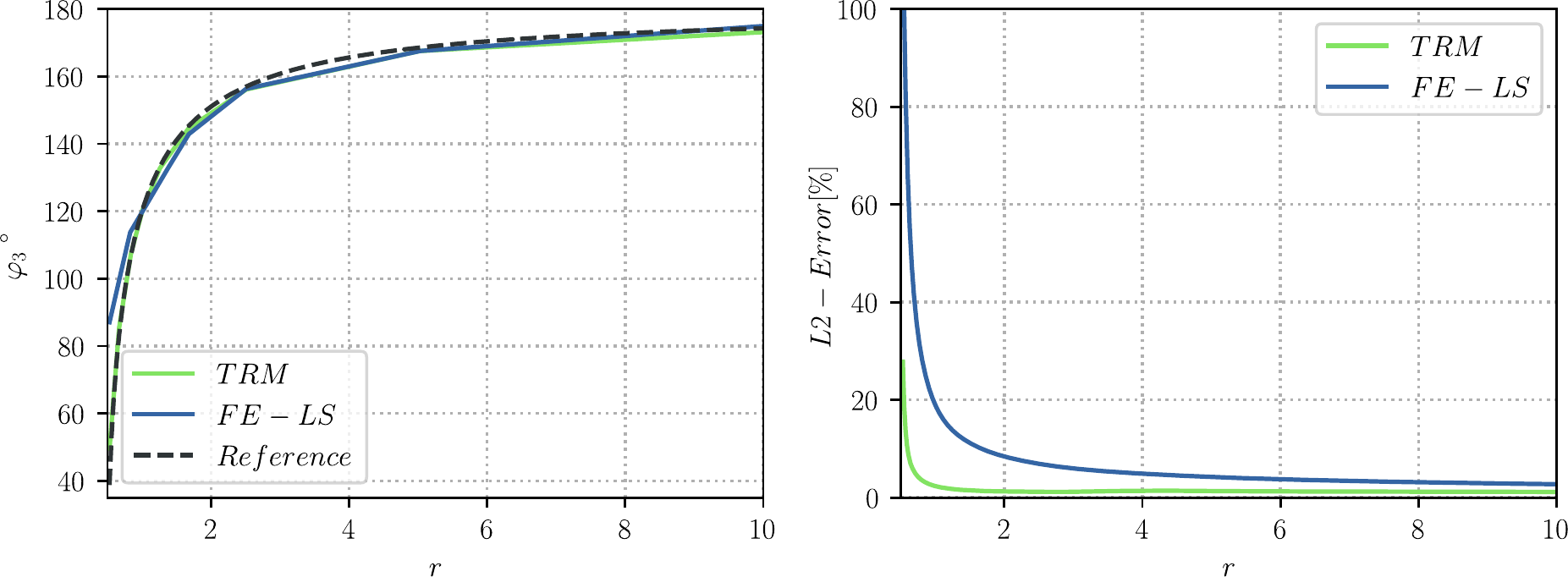}
\caption{Triple junction test case: (left) Final value for the $\varphi_3$ angle plotted against the grain boundary energy ratio $r$ and compared to the analytical equilibrium value, and (right) L2-Error. The plotted data for the LS-FE model was taken from \cite{Fausty2018}.}
\label{fig:TriplePointFinalAll}
\end{figure}

Similarly to \cite{Fausty2018}, we tested ratios in the range of $r = [0.53, 10]$, and the obtained equilibrium angles were compared to the analytical equilibrium state obtained thanks to Eq. \ref{Eq:YoungsEquilibriumAngles}. Fig.~\ref{fig:TriplePointAngleEvolution}(left) illustrates the evolution of the $\varphi_3$ angle for different values of r obtained with the TRM model. These values are compared to the ones obtained in \cite{Fausty2018} (see Fig.~\ref{fig:TriplePointAngleEvolution}(right)), where we have found that the TRM model evolves faster to its equilibrium state than the \emph{heterogeneous} FE-LS method for values of $r>1.67$. Also, the TRM model can reproduce more accurately the analytical values of $\varphi_3$ for $r<1.0$. Fig.~\ref{fig:TriplePointFinalAll} also illustrates this aspect, where the final value of $\varphi_3$ is plotted against the grain boundary energy ratio $r$ and compared to the analytical equilibrium value via an L2-Error plot.\\

\begin{figure}[!h]
\centering
\includegraphics[width=1.0\textwidth] {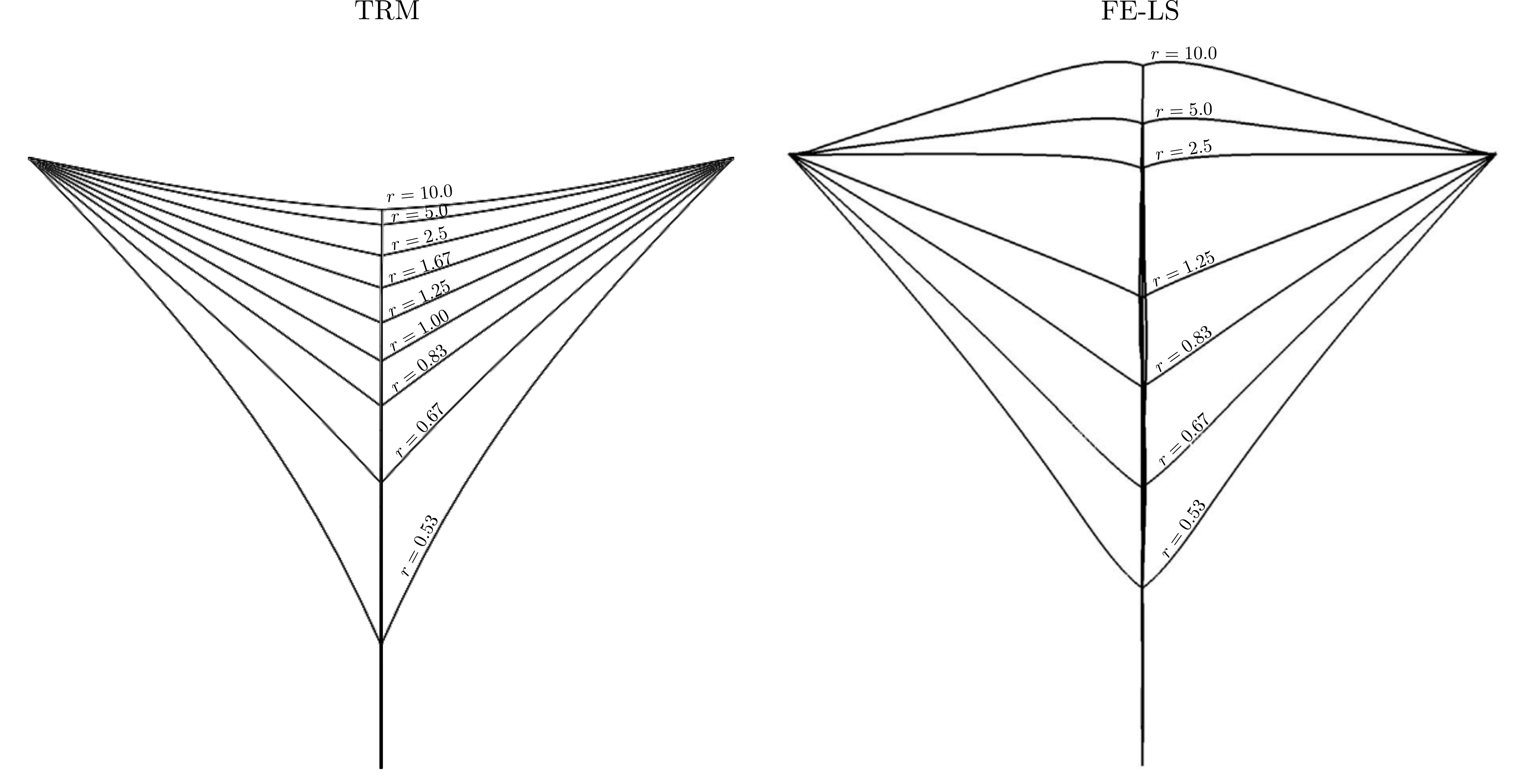}
\caption{Triple junction test case: final interface states for the (left) TRM and (right) LS-FE models. The displayed for the LS-FE model was taken from \cite{Fausty2018}.}
\label{fig:TriplePointAllFinal}
\end{figure}

\begin{figure}[!h]
\centering
\includegraphics[width=1.0\textwidth] {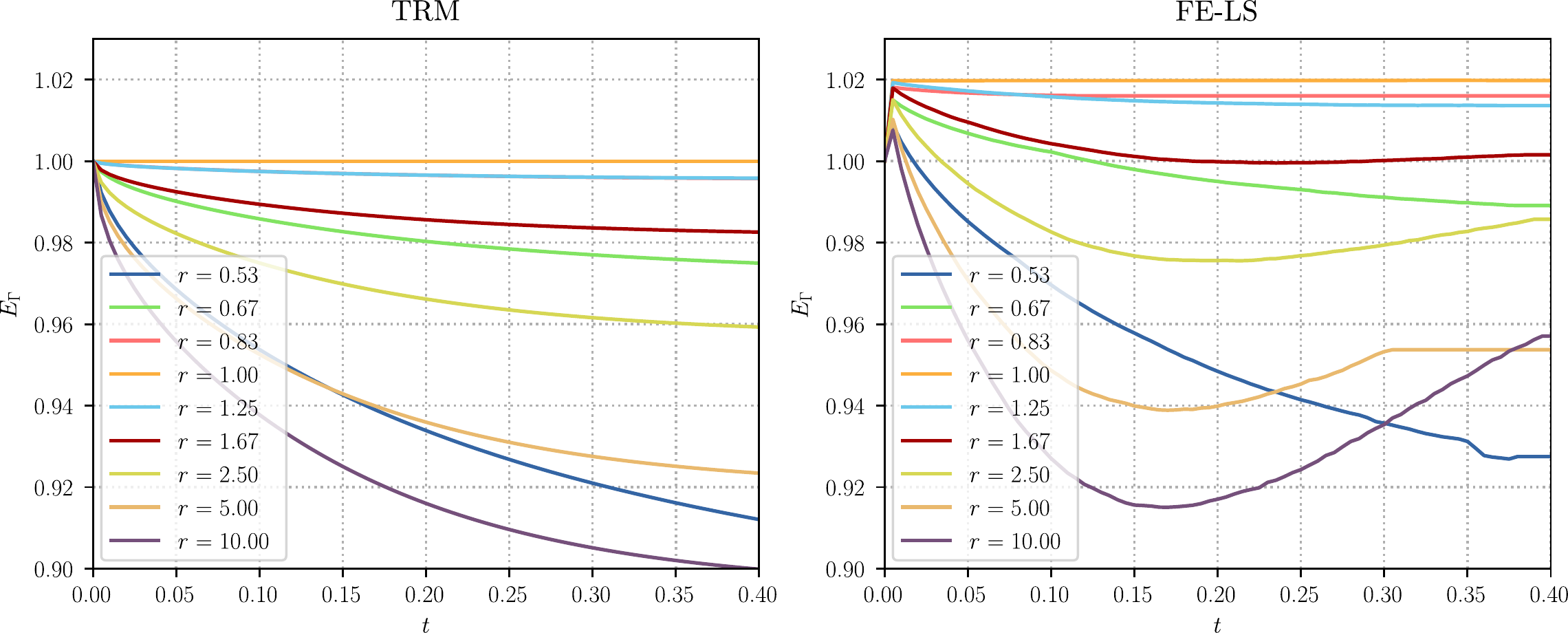}
\caption{Triple junction test case: evolution of the normalised total grain boundary energy $E_{\Gamma}$ for the (left) TRM and (right) LS-FE models. The plotted data for the LS-FE model was taken from \cite{Fausty2018}.}
\label{fig:TriplePointTotalGBE}
\end{figure}

Figure \ref{fig:TriplePointAllFinal} illustrates the final interface states for both approaches at the end of the simulation. In \cite{Fausty2018} it was found that, while the equilibrium angles of $\phi_3$ were accurately described for values of $r>2.5$ near the junction, the behaviour of the interfaces was strongly affected by the boundary conditions applied to the FE resolution, inducing non-minimal energy configurations. This behaviour was not found nor expected with the TRM model as boundary conditions only affect the velocity of nodes belonging to the boundary, and as a result, the TRM model reduces in all cases (until the equilibrium) the total energy of the system. This result can be found in Fig.~\ref{fig:TriplePointTotalGBE} where the evolution of the normalised GB energy $\overline{E}_{\Gamma}$ (each curve was scaled to start from a value equals to 1), has been plotted as a function of time.

\subsection{2D GG with heterogeneous GB properties}

Similarly to the triple junction test, in this section we reproduce the same testing approach of \cite{Fausty2020} for \emph{heterogeneous} FE-LS simulations of 2D-GG. \\

The first set of simulations measures the accuracy of the TRM model to reproduce results using different sets of mesh size and time step ($h_{trm}, dt$). Results of these simulations are given in Appendix \ref{apenSensiMeshTime}. These simulations used a Read-Shockley (RS) type function \cite{Read1950} for the determination of the GB energy $\gamma$ as a function of the disorientation angle $\theta$:

\begin{equation}
\label{Eq:ReadShocleyEquation}
\centering
\gamma=
\begin{cases}
\gamma_{max}\left(\frac{\theta}{\theta_{max}}\right)\left(1-\ln\left(\frac{\theta}{\theta_{max}}\right)\right) & \theta<\theta_{max}\\
\gamma_{max} & \theta\geq\theta_{max}
\end{cases}
\end{equation}

where $\gamma_{max}$ is the maximal grain boundary energy equals to $1.012~J m^{-2}$, $\theta_{max}$ corresponds to a threshold angle of $30^\circ$ and the mobility term $\mu$ has been held constant and equal to 0.1 $mm^4 J^{-1} s^{-1}$. These values are identical to the ones used in \cite{Fausty2020} for pure Nickel at 1400 $K$, where the authors have explained that contrary to the common usage of the RS function (using values for $\theta_{max}$ in the range of [10, 15] $^\circ$) a value of  $\theta_{max}=30^\circ$ enables to numerically increase the system's heterogeneity.

However, even with this choice, the heterogeneity level using a RS type function remains minimal. Indeed, only a narrow percent of the grain boundaries present a disorientation angle in the "variational" zone of the RS function (see Fig.~\ref{fig:2DGGHetero_5000_InitialState}(bottom-right) for the values with a disorientation angle $\theta<30^\circ$.) while the majority of the interfaces present a disorientation angle $\theta\geq 30^\circ$, and thus they acquire a value of $\gamma=\gamma_{max}$. In \cite{Fausty2020}, as a means to extend the representativity of the \emph{heterogeneous} LS-FE formulation, multiple functions were used to compute the value of $\gamma$ as a function of $\theta$. This section will test the TRM model using two of the proposed functions. As such, the results presented here can be directly compared to those detailed in \cite{Fausty2020}. These functions correspond to the \emph{Read-Shockley+} and the \emph{Gaussian} functions, which produce the most heterogeneous configurations. These functions are defined as follows:

{\bf RS+}
\begin{equation}
\label{Eq:ReadShocleyPlusEquation}
\centering
\gamma=
\begin{cases}
\gamma_{max}'\left(\frac{\theta}{\theta_{max}}\right)\left(1-\ln\left(\frac{\theta}{\theta_{max}}\right)\right) & \theta<\theta_{max}\\
\gamma_{max}' & \theta_{max}\leq\theta\leq\theta_{thresh}\\
0.1\gamma_{max}' & \theta>\theta_{thresh}
\end{cases}
\end{equation}
where $\gamma_{max}'=1.1~Jm^{-2}$ and $\theta_{thresh}=55^\circ$

{\bf Gaussian}
\begin{equation}
\label{Eq:ReadShocleyPlusEquation}
\centering
\gamma= \gamma_g \exp{\frac{-(\theta-\theta_{\mu})^2}{2\theta_{\sigma}^2}}
\end{equation}
where $\gamma_{q}'=1.54~Jm^{-2}$, $\theta_{\mu}=40^\circ$ and $\theta_{\sigma}=10^\circ$

\begin{figure}[!h]
\centering
\includegraphics[width=1.0\textwidth] {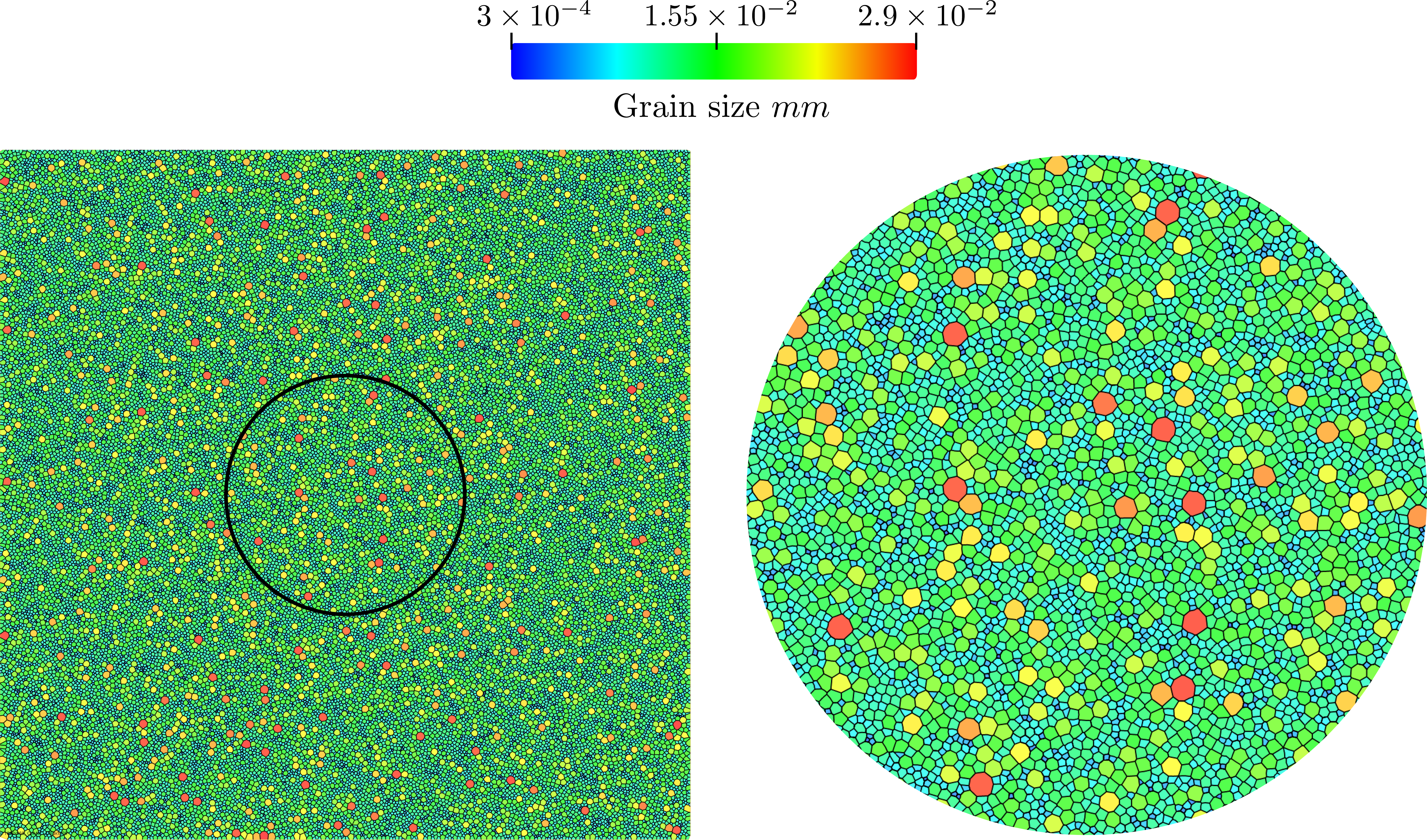}
\caption{Example of the initial state of a 2D heterogeneous GG test case with 40000 initial grains, (left) grain size field, (right) zoom of the circular section.}
\label{fig:2DGGHetero_50000_InitialState}
\end{figure}

\begin{figure}[!h]
\centering
\includegraphics[width=1.0\textwidth] {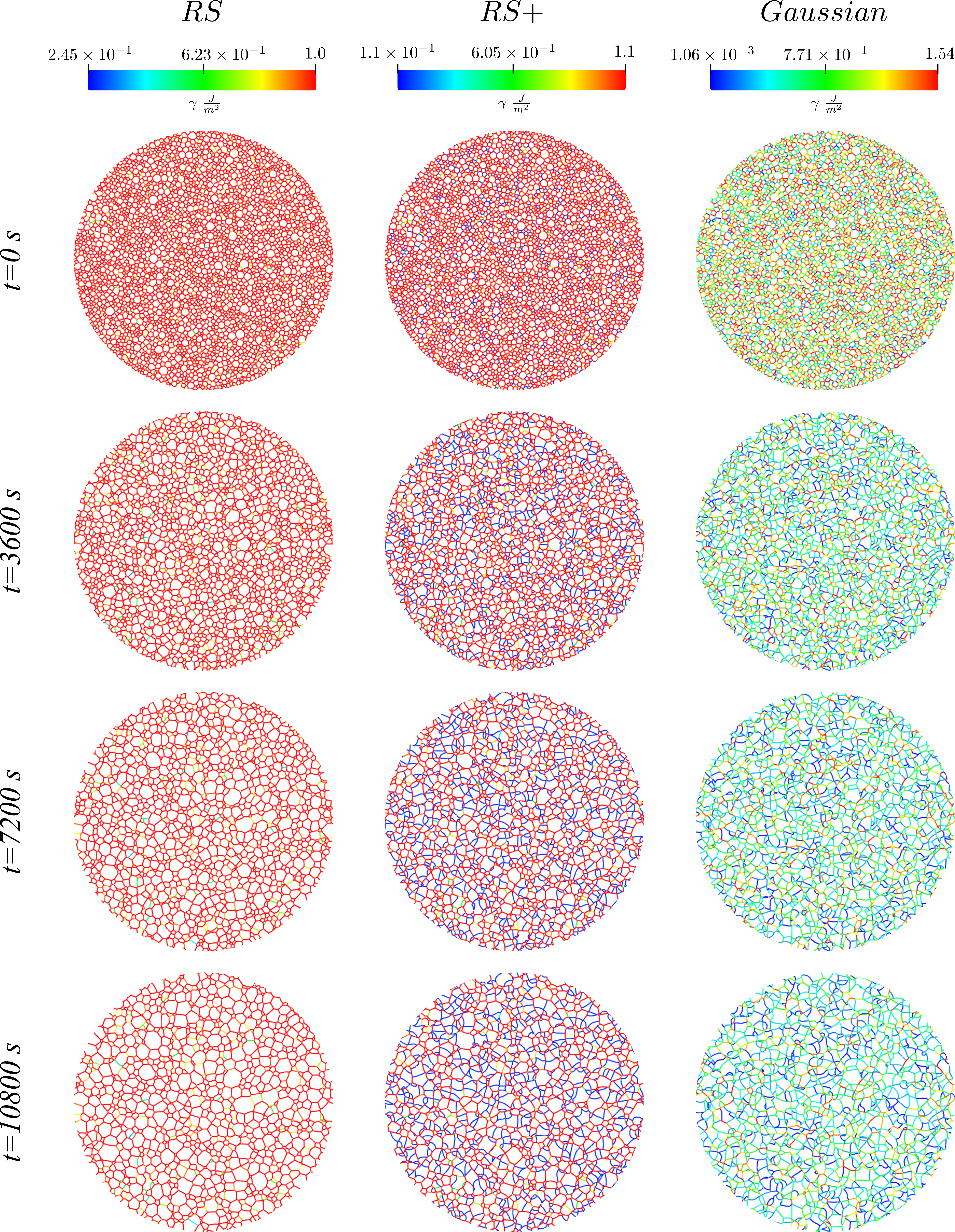}
\caption{Examples of the evolution of the microstructure, from top to bottom as a function of time, and using the functions from right to left: RS,  RS+, and Gaussian. A much higher heterogeneity is found in the cases using the Gaussian and RS+ functions. These figures illustrate the presence of stable high-order multiple junctions.}
\label{fig:States_5000}
\end{figure}

These formulations were used along with the RS function and a homogeneous formulation ($\gamma=1.012~Jm^{-2}$) in the TRM model for the full-field modelling of annealing. These simulations were performed over four different initial Laguerre-Voronoi tessellations \cite{Hitti2012} based on an optimized sphere packing algorithm \cite{hitti2013optimized} and representative of the same statistical grain size distribution given in \cite{Fausty2020} with approximately 40000 grains each. One example of initial tessellation is given in Fig.~\ref{fig:2DGGHetero_50000_InitialState}. Hereafter, all results will contain data taken from the results of the four initial states and mean quantities will be averaged. Fig.~\ref{fig:States_5000} shows the evolution of this tessellation in time and for different $\gamma$ functions, here, the colours are representative of the GB energy of each interface. These figures illustrate how the RS formulation is too ``homogeneous", presenting just a few variations in the GB properties (even at the end of the simulation), while the RS+ and Gaussian are more heterogeneous. Additionally, in the RS+ and Gaussian cases, interfaces with a high GB energy seem to be eliminated during the early stages of the simulations, giving a higher predominance to low-energy GBs, which is not the case for the RS configuration. Another essential aspect observed in the final states of the RS+ and Gaussian cases is the appearance of stable high-order multiple junctions as predicted in section \ref{sec:highOrderMJSplitting}.\\

\begin{figure}[!h]
\centering
\includegraphics[width=1.0\textwidth] {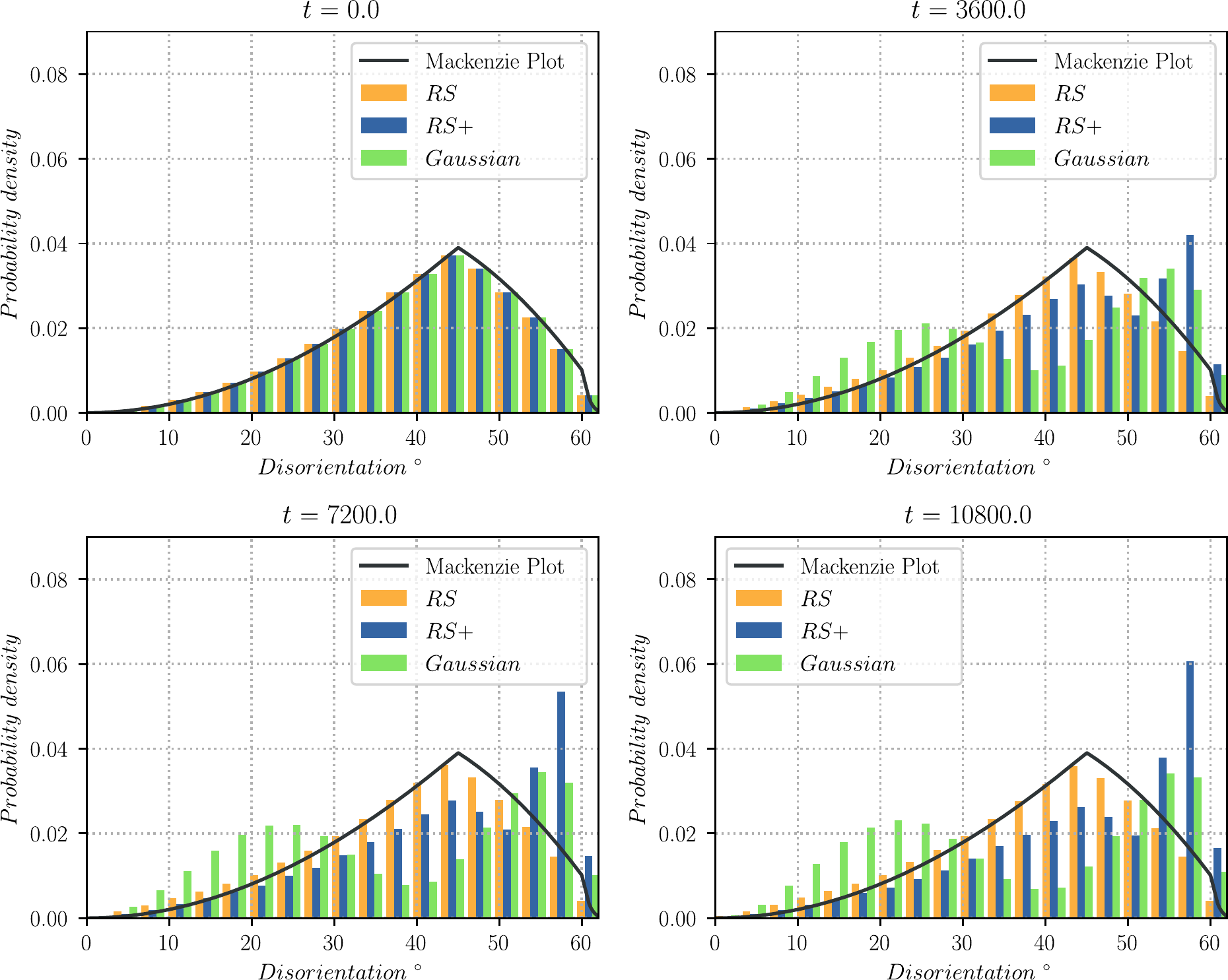}
\caption{Probability density plot of the disorientation angle weighted by GB length, for the cases using the Gaussian, RS, and RS+ functions for the computation of the GB energy $\gamma$, and for different times. Each class of data contains the results of four different initial states representative of the same initial grain size distribution.}
\label{fig:DisorientationProbabilityDensity_5000}
\end{figure}

\begin{figure}[!h]
\centering
\includegraphics[width=1.0\textwidth] {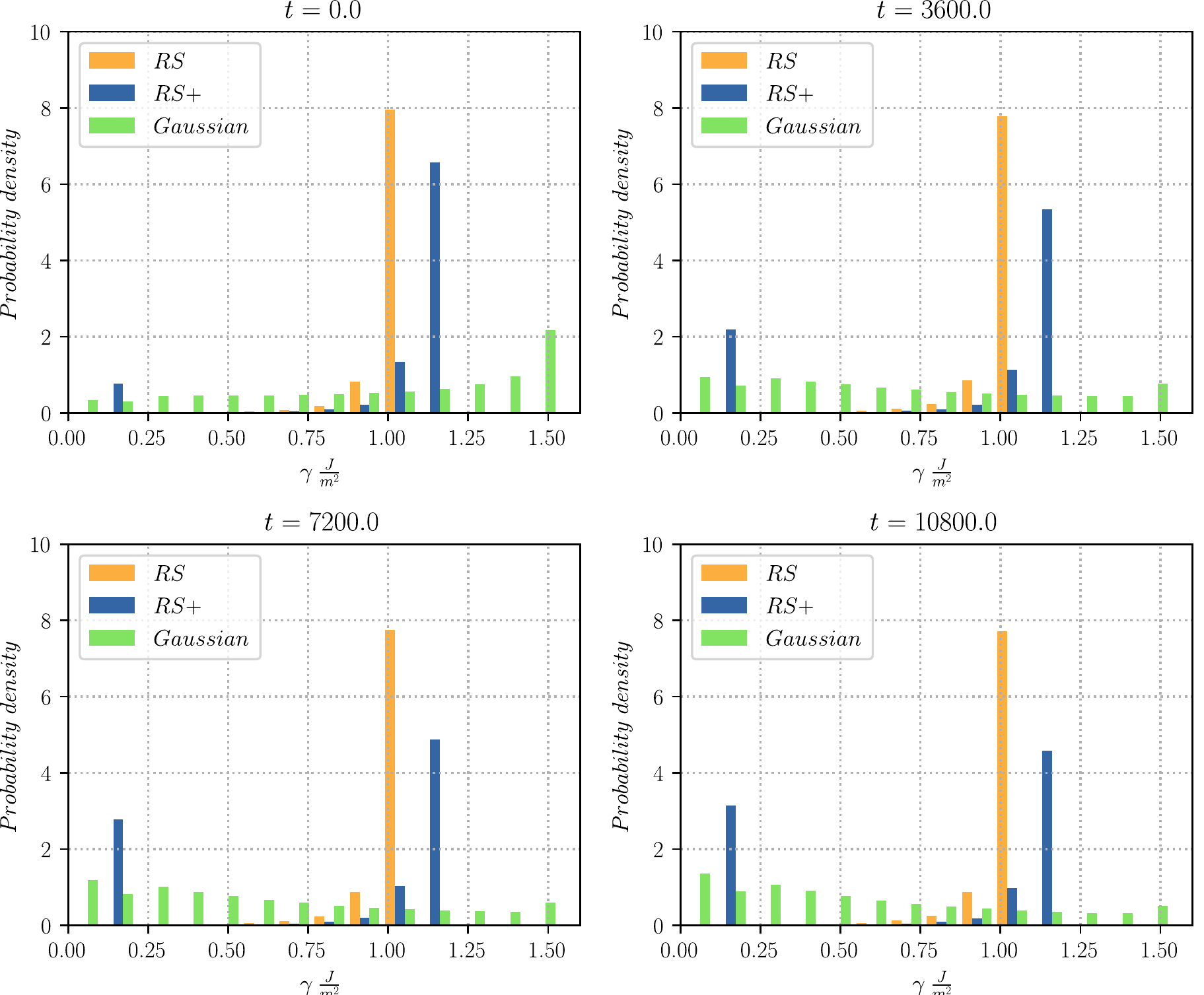}
\caption{Normalised grain boundary energy distribution using various functions for the computation of the GB energy $\gamma$. The distributions are given for every hour of thermal treatment. Each class of data contains the results of four different initial states representative of the same initial grain size distribution.}
\label{fig:EnergyProbabilityDensity_5000}
\end{figure}

Figure \ref{fig:DisorientationProbabilityDensity_5000} illustrates the normalised GB disorientation distributions of the heterogeneous configurations for every hour of annealing. Results show how the RS maintains its shape near the Mackenzie plot, hence not giving almost any preference to low energy GBs. Contrarily, the Gaussian and RS+ cases tend to avoid the disorientation angles with high energy. The plot shows maximum values at disorientation angles with low energy (e.g., for the RS+ configuration, finds one maximum at $\theta>\theta_{thresh}=55^\circ$). These results can also be observed in terms of the normalised grain boundary energy distributions given in Fig.~\ref{fig:EnergyProbabilityDensity_5000}. In only one hour of annealing, the Gaussian and RS+ configurations tend to dissipate high energy GBs, giving a much higher predominance to low energy GBs and promoting their permanency (or their appearance) as time advances. In contrast, for the RS configuration, the changes in the distribution of energy remain negligible. The Gaussian configuration is a perfect example of how the TRM algorithm respond to grain boundary energy minimisation when opposed to a highly heterogeneous configuration.\\

\begin{figure}[!h]
\centering
\includegraphics[width=1.0\textwidth] {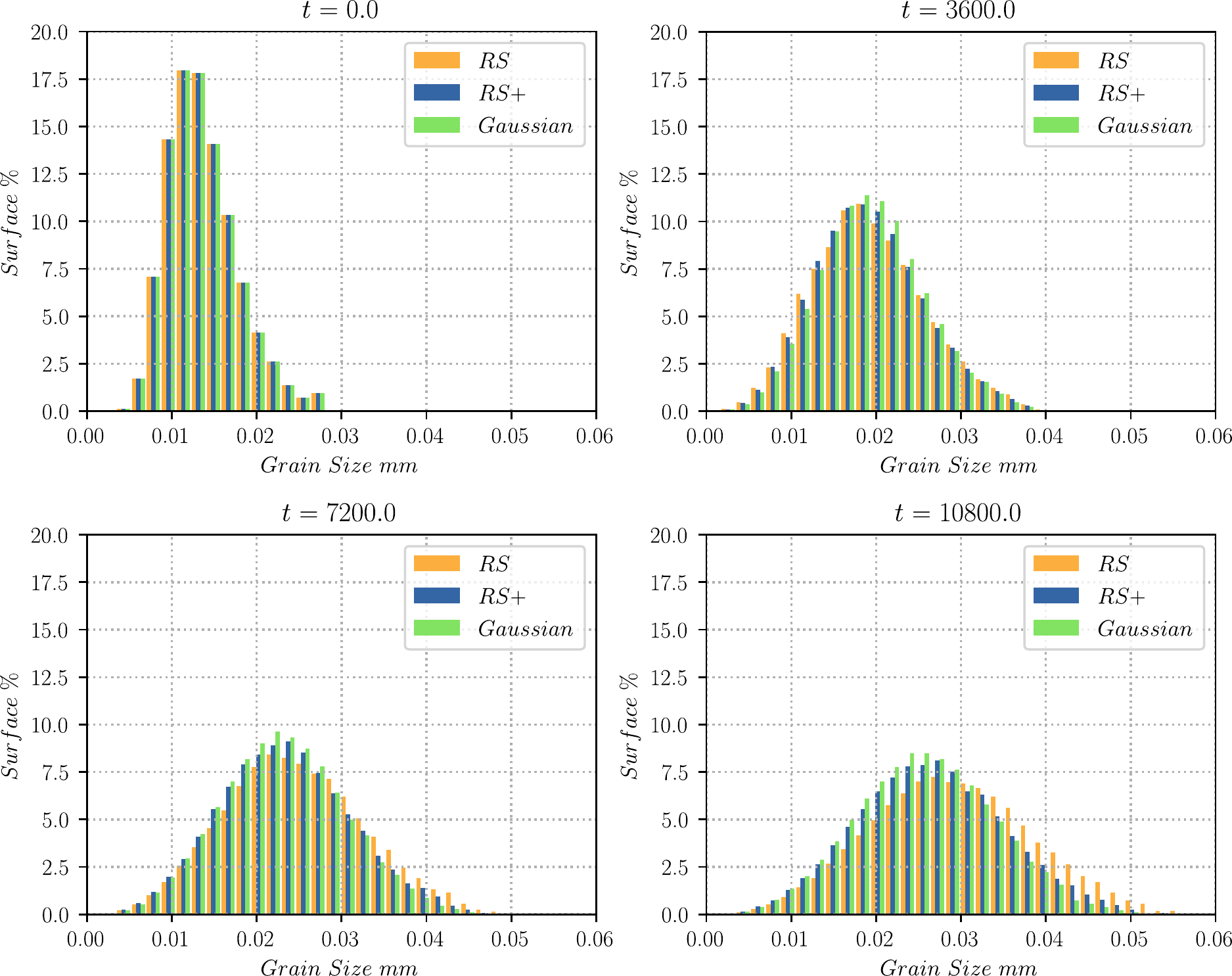}
\caption{Grain size distributions weighted by surface using various functions for the computation of the GB energy $\gamma$. The distributions are given for every hour of thermal treatment. Each class of data contains the results of four different initial states representative of the same initial grain size distribution.}
\label{fig:GrainSizeDistribution_5000}
\end{figure}

Low energy GBs predominance may produce a deceleration of the evolution of the grain size in the domain. Fig.~\ref{fig:GrainSizeDistribution_5000} illustrates the grain size distribution of the different test cases showing how the RS configuration produces a grain size distribution with larger sizes while the RS+ and Gaussian configurations promote smaller grains.\\

\begin{figure}[!h]
\centering
\includegraphics[width=1.0\textwidth] {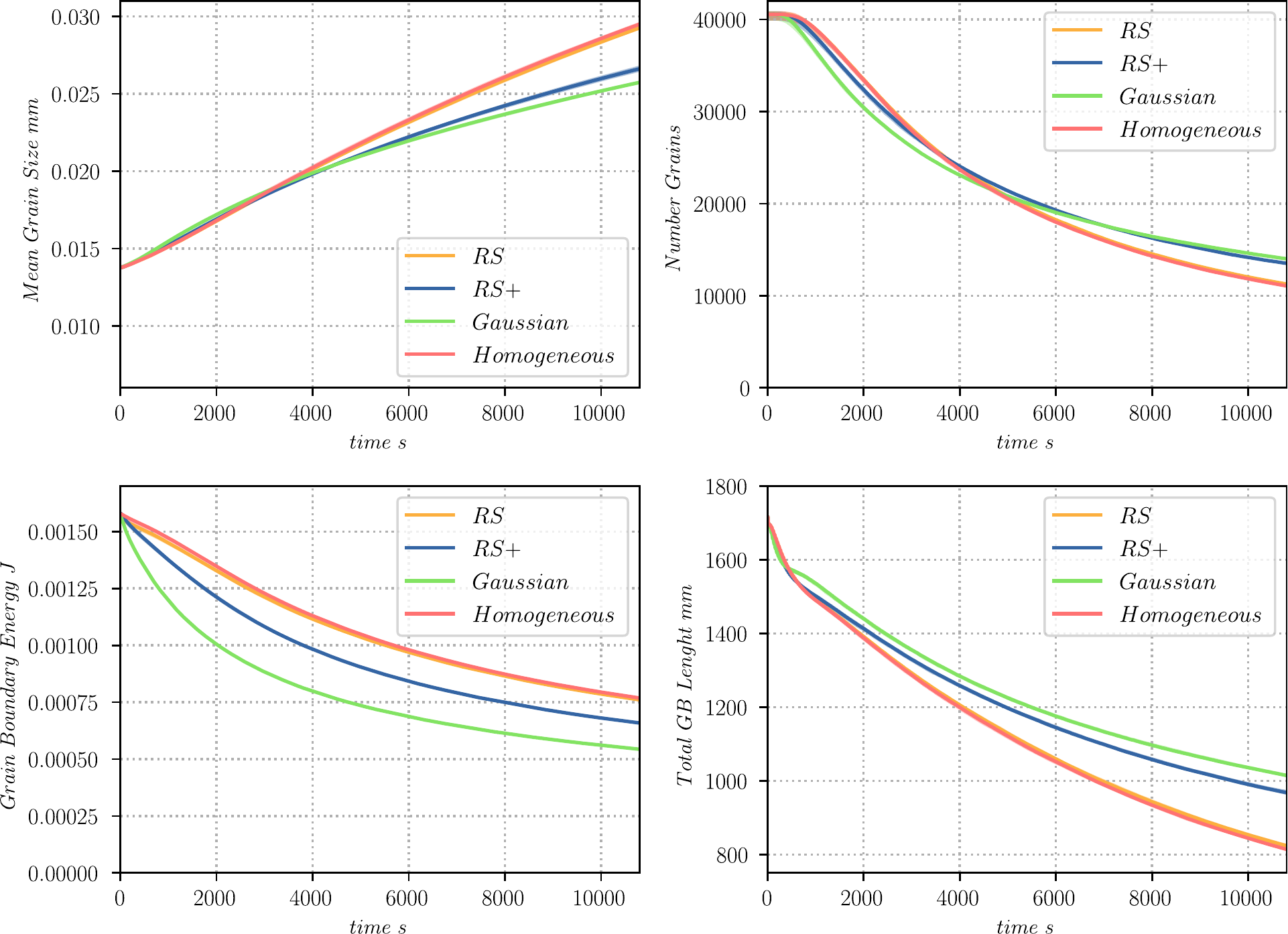}
\caption{Evolution of different parameters as a function of time, for the 2D heterogeneous GG test case simulated with the TRM model: (top-left) mean grain size pondered by surface, (top-right) number of grains, (bottom-left) total grain boundary energy $E_{\Gamma}$ and (bottom-right) CPU-time of the simulation}
\label{fig:EvolutionParameters_5000}
\end{figure}

Fig.~\ref{fig:EvolutionParameters_5000} gives the evolution of the mean grain size, the total number of grains, the total GB energy, and the total grain boundary length. The minimisation of the GB energy is much higher for the most heterogeneous cases (RS+ and Gaussian), even though their number of grains and mean size appears to have a "slower" evolution than the RS and homogeneous cases. Also, the responses of the homogeneous and the RS cases are very similar. 

\begin{figure}[!h]
\centering
\includegraphics[width=1.0\textwidth] {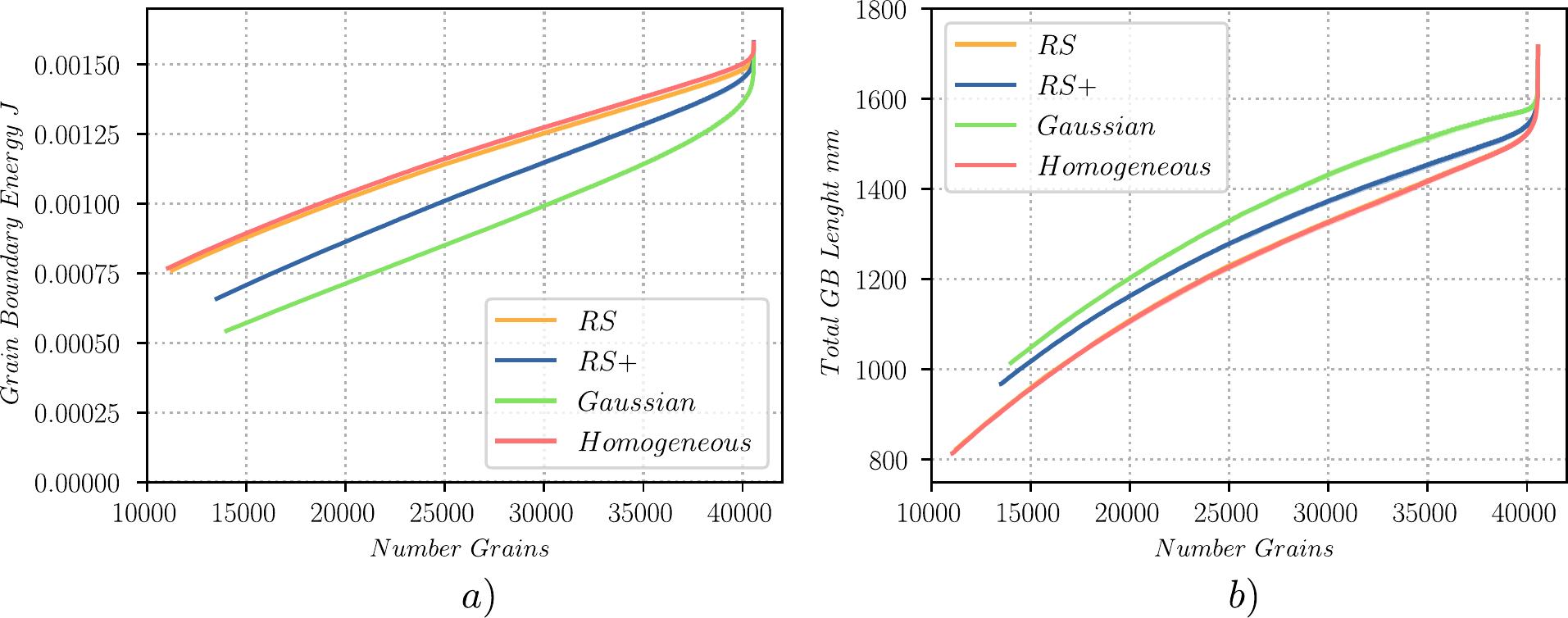}
\caption{Computational cost of the 2D heterogeneous GG test case as a function of a) the number of grains and c) the total GBs length. Additionally, b) gives the total length of GBs as a function of the number of grains. The red circles in every frame give the beginning of the simulation. The heterogeneous cases present a much higher computational cost than the homogeneous case due to the computation of the misorientation and disorientation angles at the interfaces.}
\label{fig:CPU-TimePer-Increment}
\end{figure}

Fig.~\ref{fig:CPU-TimePer-Increment}(b) gives the total GB length plotted against the number of grains of the simulation showing how the Gaussian and RS+ cases have a higher value than the RS and the homogeneous cases. This result can not be anticipated as one could have guessed the contrary, by seeing the evolution curves of the total GB energy given in Fig.~\ref{fig:EvolutionParameters_5000}(bottom-left) as a function of time and as illustrated in Fig.~\ref{fig:CPU-TimePer-Increment}(a) as a function of the number of grains. This result is a product of the preference of the higher heterogeneous cases for grain boundaries of low energy, but also by the more frequent apparition of high-order multiple junctions that decelerate the reduction of the total GB length.\\

Table~\ref{fig:EvolutionParameters_5000} gives the CPU-time of each simulation, showing how the computational cost of the TRM model in this context may be more related to the length of boundaries than to the number of grains (see Fig. \ref{fig:CPU-TimePer-Increment}b)). Additionally,  the differences between the computational cost of the homogeneous and the heterogeneous cases are very high. This can be explained by the fact that for the homogeneous case, it is not necessary to compute the misorientation at GBs nor the lowest energy configuration in the event of a separation of MJs. These operations are very demanding as both rely on an iterative computation of the lowest rotation angle between two orientations in a set of 24 possible rotations, where all of them have to be tested. Table ~\ref{fig:EvolutionParameters_5000} also shows that, the higher the heterogeneity of the case, the higher is its computational cost. This behaviour can be anticipated by seeing the evolution of the number of grains and the total length of boundaries (Fig.~\ref{fig:CPU-TimePer-Increment}), as the more homogeneous cases reduce these quantities much faster.\\

\begin{table}[h]

\caption{CPU-time of all simulations. Mean values are averaged for the same $\gamma$ function and all initial states (IS).}

\begin{center}
\begin{tabular}{ |c|c|c|c|c|c| } 
 \hline
$\gamma$ function & IS 1 & IS 2 & IS 3 & IS 4 & mean\\
 \hline
$Homogeneous$ & 0h56m16s & 0h55m16s & 0h57m53s  & 0h58m11s & 0h56m54s\\
$RS$ & 7h5m48s & 6h58m36s & 7h5m15s & 7h0m17s & 7h2m29s\\
$RS+$ & 7h30m8s & 7h19m21s & 7h25m35s & 7h21m25s & 7h24m7s\\
$Gaussian$ & 7h49m6s & 7h48m22s & 7h50m28s & 7h51m30s & 7h49m51s\\

\hline
\end{tabular}
\end{center}

\end{table}

The results presented  here are very similar to the ones obtained in \cite{Fausty2020} in the context of the \emph{heterogeneous} FE-LS formulation. This suggests that both methodologies are valid to predict microstructural states in a full-field context, as even though the mechanisms behind their evolution are the same, they have been modelled using a completely different numerical scheme and still produce a very similar outcome. It mush be highlighted that simulations in \cite{Fausty2020} where performed with initial states with around 5000 grains while here we performed simulations 8 times larger (for simulations using the same initial grains as in  \cite{Fausty2020} see \cite{Florez2020e}). Moreover, the computational power needed to produce these results using the \emph{heterogeneous} FE-LS may be much higher than the one needed by the TRM. The TRM model performed all \emph{sequential} simulations in less than eight hours for the heterogeneous configurations and in less than one hour for the homogeneous ones, using an AMD Ryzen 7 3600x processor. \\

\section{Discussion, conclusion and perspectives}\label{sec:conclusions}

This article has provided the necessary implementation for modelling grain growth using heterogeneous grain boundary properties with the TRM model. These implementations consist of: \textbf{i.} a numerical framework on top of the TRM base code to measure neighbors' misorientation. The algorithm only takes these measurements at grain interfaces, namely, \emph{L-Nodes} and \emph{PP-Connections}. \textbf{ii.} A decomposition algorithm for high-order multiple junctions, which searches for the lowest energy configuration among all possible decompositions. These decompositions are obtained by the separation of pairs of interfaces from the MJ, storing for each, the total GB energy change $\Delta E_{\Gamma}$ and applying the one with the lowest $\Delta E_{\Gamma}$ only if it is negative (as for events with a minimum value of $\Delta E_{\Gamma}>0$ the original configuration should remain stable). Finally, \textbf{iii.} a formulation for the computation of the velocity using anisotropic data was established using a discrete formulation inspired by the literature \cite{Kawasaki1989,  BarralesMora2010}. \\

The new methodology implemented for the TRM model was tested in the context of heterogeneous grain boundary properties, using identical test cases like the ones presented in \cite{Fausty2018, Fausty2020}. Results show that the TRM model can produce more accurate results regarding the equilibrium angles of triple junctions compared to the analytical values given by Young's equilibrium. Additionally, the TRM model ensures at all times low-energy stable configurations contrary to the \emph{heterogeneous} LS-FE model presented in \cite{Fausty2018} which may produce stable configurations with non-minimal energy states. Furthermore, the TRM model was tested in a GG context using heterogeneous grain boundary properties in function of the disorientation angle. The initial configurations of all tests were statistically identical to the one presented in \cite{Fausty2020} with around 40000 initial grains. Sensitivity analyses were performed, resulting in a tendency of the model to converge to a fixed solution when decreasing the set of parameters ($h_{trm},~dt$), controlling the mesh size and the time step, respectively. Then, multiple formulations for the determination of the grain boundary energy $\gamma$ as a function of the disorientation angle $\theta$ were used, namely the Read-Shockley (RS) \cite{Read1950}, the modified Read-Shockley (RS+), and the Gaussian formulations used in \cite{Fausty2020}. Results showed a similar statistical behaviour to the results presented in \cite{Fausty2020} in a LS-FE context, hence validating both approaches at this scale.\\

Results also show that the CPU-time depends on to the total length of GBs. Additionally, the computational needs of the heterogeneous cases are higher than when using a homogeneous configuration. This result is strongly related to the computation of the disorientation angle which, even if it is only performed at the interfaces, it remains a brute force algorithm, which in \cite{Florez2020e} showed a poor performance, taking up to 60\% of the total CPU-time in the heterogeneous configurations.\\

Finally, it remains a perspective of the present work to test the TRM model in a fully anisotropic environment, where the influence of the inclination of the interface on the computation of $\gamma$ is taken into account, producing variations of properties over curved GB and torque terms. Such a study will be presented in a forthcoming publication \cite{Florez2021Statistical}.

\appendix

\section{Appendices}
\subsection{Algorithm for the decomposition of high-order multiple junctions}\label{apenDecompositionAlgorithm}

Algorithm \ref{alg:DecomposeAniso} summarizes the TRM implementation of MJ decomposition. Here we have used the function $E_b(B)$ which gives the total surface energy of the internal boundary segments $B$, of a given element patch $e_p$, obtained thanks to the function $Boundaries(e_p)$, also, we use the function $GL(i,j,N)$ which returns a list of size $i$ with the $j_{th}$ set of consecutive\footnote{Consecutiveness is measured in this context following polar coordinates (i.e. the angle made by a given line and the $x$ axis)} boundary segments attached to Node $N$ (i.e. for the case given in Fig. \ref{fig:PossibleDecompositions}, $GL_{(2,1, N_j)}=\{\overline{N_jN_3}, \overline{N_jN_4}\}$, $GL_{(2,2, N_j)}=\{\overline{N_jN_4}, \overline{N_jN_5}\}$, $GL_{(2,3, N_j)}=\{\overline{N_jN_5}, \overline{N_jN_1}\}$, $GL_{(3,1, N_j)}=\{\overline{N_jN_3}, \overline{N_jN_4}, \overline{N_jN_5}\}$... ) and finally, the function $Boundaries(e_{p})$. The algorithm first searches between each pair of consecutive segments, the one that would reduce the boundary energy the most if it is separated from the MJ (just as depicted in Fig.~\ref{fig:PossibleDecompositions}), and selects this configuration. Then, if this configuration reduces the initial GB energy given by the initial state, the initial configuration is replaced, and the algorithm continues to the next MJ. If not, instead of searching between pairs, the algorithm will re-iterate between consecutive triplets of lines if the order $z$ of the MJ is sufficiently high (at least $z=6$) and so on. Finally, if no configuration tested has lower energy than the initial configuration, the algorithm considers the MJ as stable and continues to the next. Note that the decompositions are made one at a time for a given call of algorithm \ref{alg:DecomposeAniso} over a given MJ. This means that a MJ of order $z=5$ might entirely decompose in two increments and one with $z=6$ in three.\\

This procedure, however, accepts configurations with higher energy than the \emph{total} minimal energy state (the configuration with the minimum possible boundary energy), especially for MJs of high order ($z>6$). However, in practice, such configurations have a very low probability of appearance in real microstructures.

\begin{algorithm}
\caption{MJ decomposition algorithm for the TRM model}\label{alg:DecomposeAniso}
\begin{algorithmic}[1]

\ForAll{Points: $P_i$}
	\If{$z_{(P_i)}>3$}
		\State $N_i$ $\gets$ Node representing $P_i$
		\State $e_{p0}$ $\gets$ Elements($N_i$)
		\State $B_0$ $\gets$ $Boundaries(e_{p0})$
		\State $E_0$ $\gets$ $E_b(B_0)$
		\State $E_{min}$ $\gets$ $\infty$
		\State $S_0$ $\gets$ tuple $(e_{p0},B_0)$
		\State $i$ $\gets$ 2
		
		\ForAll{ number of connections of $N_i$ : $j$}
			\State $\{L_j\}$ $\gets$ $GL_{(i,j, N_i)}$
			\State Separate $\{L_j\}$ from $N_i$ by adding a new Node $N_j$
			\State Create new boundary (PP-Connection) $\overline{N_iN_j}$ 
			
			\State $B_{j}$ $\gets$ $\overline{N_iN_j}$ $\cup$ $B_0$
			\If{$E_b(B_{j})<E_{min}$}
				\State $E_{min}$ $\gets$ $E_b(B_{j})$
				\State $e_{pj}$ $\gets$ Elements($N_j$) $\cup$ $e_{p0}$
				\State $S_{min}$ $\gets$ tuple $(e_{pj},B_{j})$
			\EndIf
		\EndFor
			\If{$E_{min}<E_0$}
				\State Replace $S_0$ by $S_{min}$
			\ElsIf{$i<z_{(P_i)}/2$}
				\State $i$ $\gets$ $i+1$
				\State {\bf goto} 10:
			\EndIf

	\EndIf
\EndFor

\end{algorithmic}
\end{algorithm}

\subsection{Computation of the disorientation angle}\label{apenMisorientComp}

We will compute the misorientation and the disorientation angle similarly as in \cite{Fausty2020}. Two neighbours grains $G_l$ and $G_m$ with orientations $O_l$ and $O_m$, respectively, form a misorientation expressed as: 

\begin{equation}
\label{Eq:MisorientationSingle}
\centering
M_{lw}^*= O_l^{-1}O_w
\end{equation}

It is, however, necessary to compute a misorientation taking into account the minimisation of the disorientation angle $\theta(O_l, O_w)$. Hence for all possible symmetric representations of the misorientation $M_{lw}^*$, with $(S_i, S_j)\in\mathbb{H}^2$ the space group of the crystal:

\begin{equation}
\label{Eq:MisorientationMinimal}
\centering
M_{lw}= \underset{\min_{i,j}\theta(O_lS_i, O_wS_j)}{S_i^{-1}M_{lw}^*S_j}
\end{equation}

The search for minimal disorientation uses a brute force algorithm. Every misorientation computation needs to iterate over all possible symmetric representations and select the one with the lowest $\theta$. In this work, we will consider only cubic-type crystals, hence 24 symmetric representations must be iterated.\\

TRM model performs these operations during two stages of the algorithm:

First, a misorientation computation is held before the computation of the nodes' velocities $v_i$, as all boundary properties must be defined at this stage. The computation is done once per \emph{Line}, which attributes all misorientations and disorientation angles for all L-Nodes and segments of the \emph{Lines} entities. Note that a \emph{Line} can only compute one misorientation, hence, it is unnecessary to compute it for every node lying on the \emph{Line}. Then, some edges still need to define their orientation: the ones describing a \emph{PP-Connection}, namely, the edges defining a connection between two \emph{Point} entities (see \cite{Florez2020b} for more information about the data structure used by the TRM model). This computation is necessary, as, even though the notion of grain boundary energy does not hold at MJs in the same way as for normal boundaries, the GB properties of all interfaces attached to the MJ are needed.\\

Secondly, the TRM model performs a misorientation computation during the decomposition of MJs for all \emph{possible new} interfaces (see line 13-15: of algorithm \ref{alg:DecomposeAniso}). This could be a very demanding procedure as, for instance, each possible decomposition seeks the minimal disorientation angle among all 24 equivalent symmetries defined for the crystal. We study the relative cost of the misorientation computation at the end of section \ref{sec:numericalresults_5}.

\subsection{Sensitivity analysis on GG simulations: mesh size and time step} \label{apenSensiMeshTime}

\begin{figure}[!h]
\centering
\includegraphics[width=1.0\textwidth] {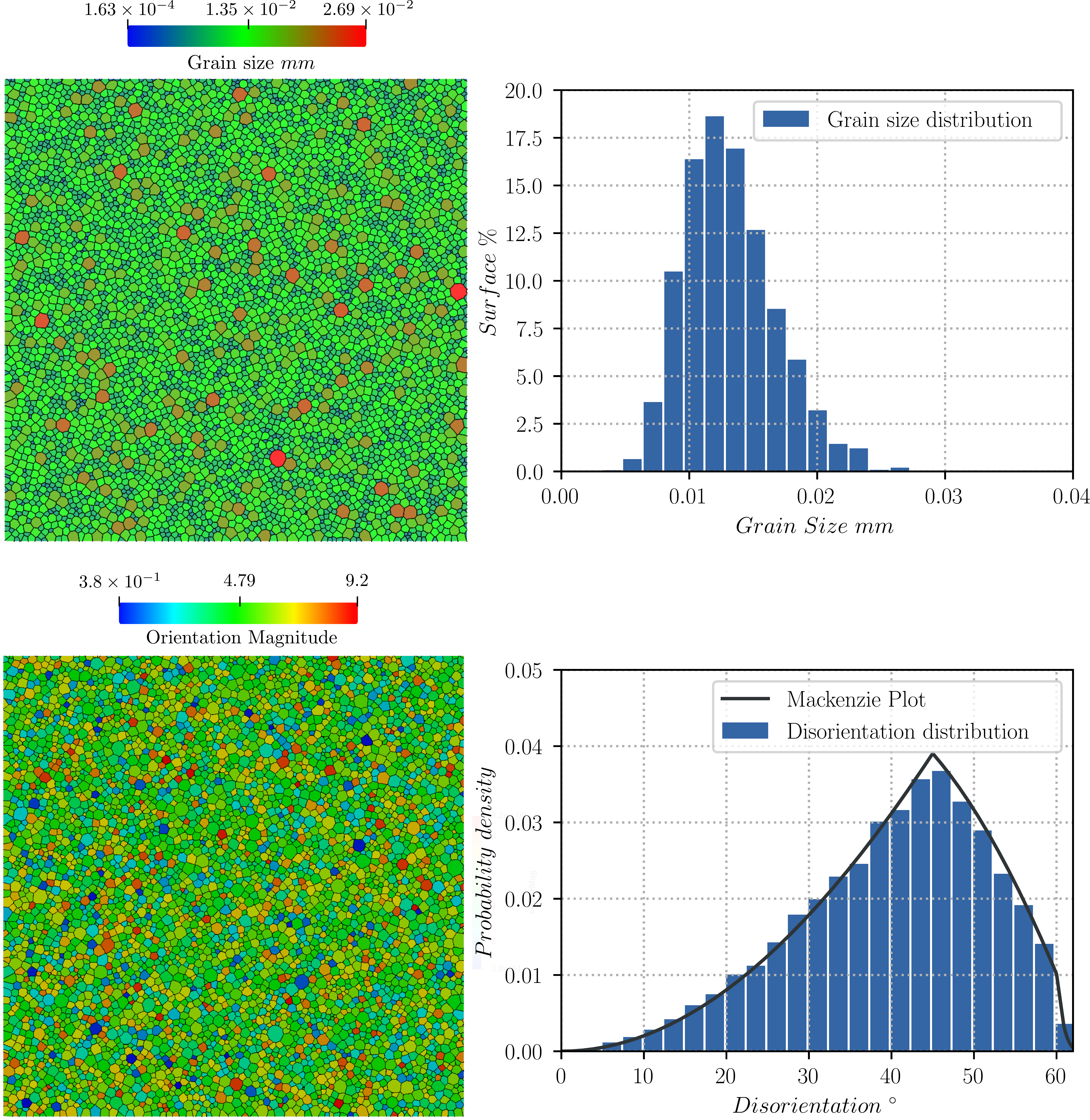}
\caption{Initial state of the 2D heterogeneous GG test case with 5000 initial grains, (top-left) grain size field, (top-right) grain size distribution weighted by surface area, (bottom-left) grain orientation field and (bottom-right) probability density plot of the disorientation angle weighted by length of interface.}
\label{fig:2DGGHetero_5000_InitialState}
\end{figure}

This section uses a squared RVE domain of $1.5~mm$ of side length to model annealing. Fig.~\ref{fig:2DGGHetero_5000_InitialState}(top-left) illustrates the initial state of the polycrystal used in this sensitivity analysis. This polycrystal contains exactly 5089 initial grains and its grain size distribution (pondered by surface) is given in Fig.~\ref{fig:2DGGHetero_5000_InitialState}(top-right). Additionally, in all cases, the mobility term has been held constant and equal to 0.1 $mm^4 J^{-1} s^{-1}$.\\

Fig.~\ref{fig:2DGGHetero_5000_InitialState}(bottom-left) shows the initial microstructure colored following the \emph{orientation magnitude} $e=\sqrt{\varphi_1^2+\Phi^2+\varphi_2^2}$~of each grain. Disorientation angles ($\theta$) have been computed for each \emph{Line}, \emph{L-Node} and \emph{PP-Connection} (i.e. for all nodes and segments belonging to the GBs) of the interface using the methodology presented in Appendix \ref{apenMisorientComp}. Fig.~\ref{fig:2DGGHetero_5000_InitialState}(bottom-right) gives the disorientation angle distribution of the initial microstructure, which shows a good agreement with the Mackenzie plot.\\

\begin{figure}[!h]
\centering
\includegraphics[width=1.0\textwidth] {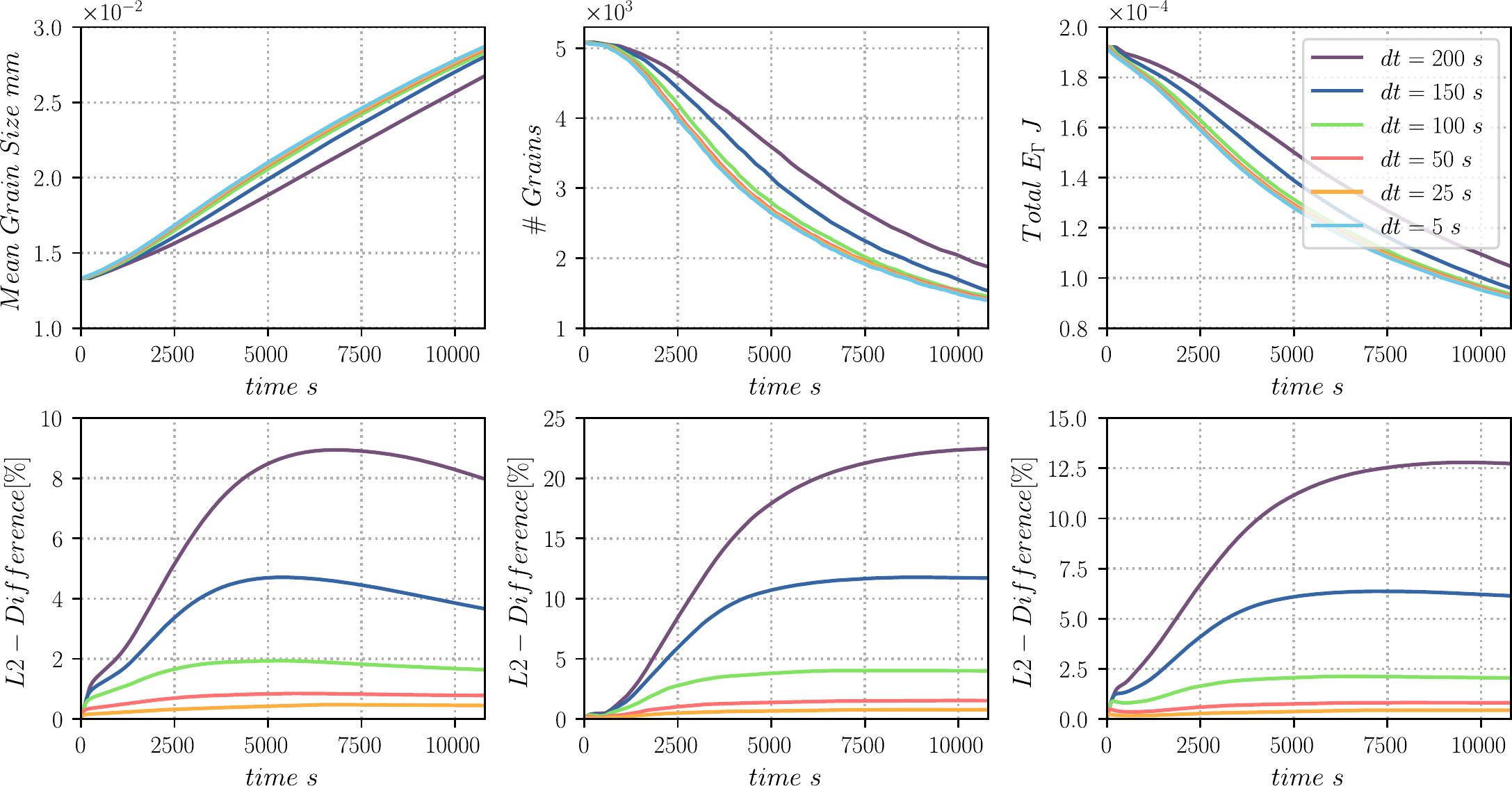}
\caption{Sensitivity to the time step $dt$ for the 2D heterogeneous GG test case using the TRM model and $h_{trm}=0.003~mm$. The evolution of different parameters is given when using the RS function for the determination of the GB energy $\gamma$ as a function of the disorientation angle $\theta$: (left) grain Size, (center) number of grains, and (right) total GB energy. Each value (top) is compared to the evolution of the smallest $dt$.  L2-difference values are given (bottom).}
\label{fig:SensitivityTimeInfluence}
\end{figure}

\begin{figure}[!h]
\centering
\includegraphics[width=1.0\textwidth] {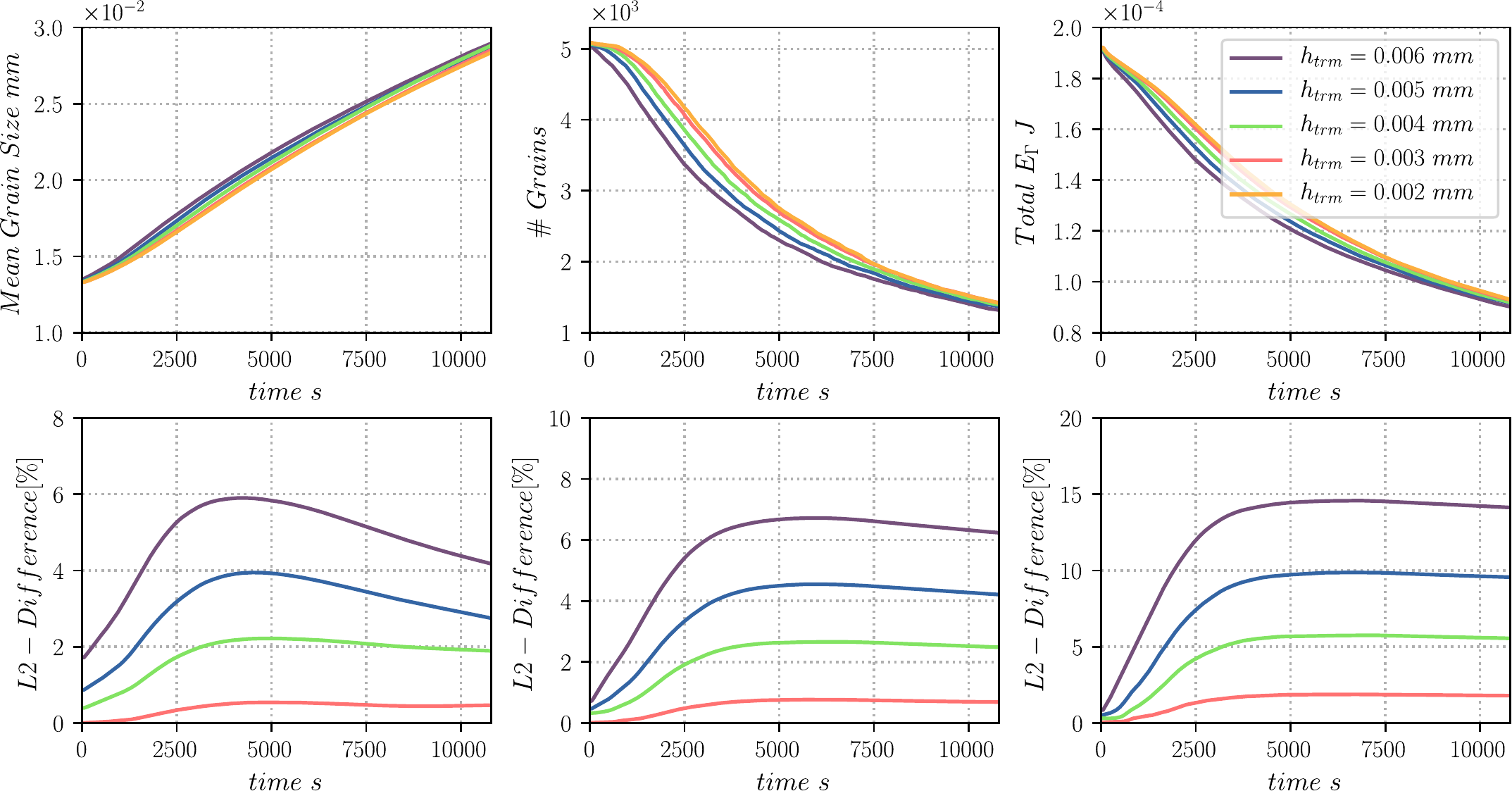}
\caption{Sensitivity to the mesh size $h_{trm}$ for the 2D heterogeneous GG test case using the TRM model and $dt=50~s$. The evolution of different parameters is given when using the RS function for the determination of the GB energy $\gamma$ as a function of the disorientation angle $\theta$: (left) grain Size, (center) number of grains, and (right) total GB energy. Each value (top) is compared to the evolution of the smallest $h_{trm}$. L2-Difference values are given (bottom).}
\label{fig:SensitivityMeshInfluence}
\end{figure}

Figure \ref{fig:SensitivityTimeInfluence}(top) gives the evolution of various parameters for the first 3 hours of simulated time using a constant mesh size of $h_{trm}=0.003~mm$ and for various time steps $dt$. The mean grain size, the number of grains, and the total GB energy have been plotted, showing a tendency to converge to a fixed solution when the time step decreases. Fig.~\ref{fig:SensitivityTimeInfluence}(bottom) gives the L2-difference of each iteration taking as a reference the curve using $dt=5~s$, confirming these results. Similarly, the simulations were repeated using a constant time step $dt=50~s$ and for various mesh sizes. Similarly as before, decreasing the mesh size produces a tendency to converge to a fixed evolution (see Fig.~\ref{fig:SensitivityMeshInfluence}), reducing the L2-difference to the reference curve (here the one using $h_{trm}=0.002~mm$) with every iteration. This study shows that one can expect good accuracy when using a set of parameters ($h_{trm},~dt$) in the surroundings of ($0.003~mm$, $50~s$). These values will be used in all other polycrystal simulations.\\

\bibliography{ms}

\end{document}